\documentclass[a4paper,12pt]{article}
\pdfoutput=1 

\usepackage{jheppub}
\usepackage{graphicx}
\usepackage{axodraw}
\usepackage{epsfig}
\usepackage{amsmath, amsthm, amssymb}
\usepackage{amsfonts}
\usepackage{subfig}
\usepackage{color}
\usepackage{xspace}
\usepackage[dvipsnames]{xcolor}
\usepackage{paralist}
\usepackage{multirow}
\usepackage{paralist}

\newcommand{\beq}{\begin{equation}}
\newcommand{\eeq}{\end{equation}}
\newcommand{\bea}{\begin{eqnarray}}
\newcommand{\eea}{\end{eqnarray}}
\newcommand{\bfig}{\begin{figure}}
\newcommand{\efig}{\end{figure}}
\newcommand{\bc}{\begin{center}}
\newcommand{\ec}{\end{center}}

\def\sq2{\sqrt{2}}
\newcommand{\PZP}{\ensuremath{Z'}\xspace}

\preprint{SMU-HEP-16-09}
\title{NLO+NLL Collider Bounds, Dirac Fermion and Scalar Dark Matter in the B-L Model }

\author[a]{Michael Klasen,}
\author[b]{Florian Lyonnet,}
\author[c]{Farinaldo S. Queiroz}

 \affiliation[a]{Institut f\"ur Theoretische Physik, Westf\"alische
 Wilhelms-Universit\"at M\"unster, Wilhelm-Klemm-Stra\ss{}e 9, D-48149
 M\"unster, Germany}

\affiliation[b]{Southern Methodist University, Dallas, TX 75275, USA}

\affiliation[c]{Particle and Astroparticle Physics Division, Max-Planck-Institut f\"ur Kernphysik, Saupfercheckweg 1, 69117 Heidelberg, Germany}

\emailAdd{michael.klasen@uni-muenster.de}
\emailAdd{flyonnet@smu.edu}
\emailAdd{farinaldo.queiroz@mpi-hd.mpg.de}

\abstract{
	Baryon and lepton numbers being accidental global symmetries of the Standard Model (SM), it is natural to promote them to local symmetries. However, to preserve anomaly freedom, only combinations of B-L are viable.
In this spirit,	we investigate possible dark matter realizations in the context of the $U(1)_{B-L}$ model: \begin{inparaenum}[i)]\item Dirac fermion with unbroken B-L; \item Dirac fermion with broken B-L; \item scalar dark matter; \item two component dark matter\end{inparaenum}. We compute the relic abundance, direct and indirect detection observables and confront them with recent results from Planck, LUX-2016, and Fermi-LAT and prospects from XENON1T. In addition to the well known LEP bound $M_{Z^{\prime}}/g_{BL} \gtrsim 7$~TeV, we include often ignored LHC bounds using 13 TeV dilepton (dimuon+dielectron) data at next-to-leading order plus next-to-leading logarithmic accuracy. We show that, for gauge couplings smaller than $0.4$, the LHC gives rise to the strongest collider limit. In particular, we find $M_{Z^{\prime}}/g_{BL} > 8.7$~TeV for $g_{BL}=0.3$. We conclude that the NLO+NLL corrections improve the dilepton bounds on the $Z^{\prime}$ mass and that both dark matter candidates are only viable in the $Z^{\prime}$ resonance region, with the parameter space for scalar dark matter being fully probed by XENON1T. Lastly, we show that one can successfully have a minimal two component dark matter model.}


\begin{document} 
\maketitle
\flushbottom

\clearpage
\section{Introduction}
\label{sec:1}

The availability of data from collider, direct and indirect searches for dark matter has raised the importance of dark matter complementarity across these search strategies. In this context, effective field theories and simplified models have become popular tools, as they can capture most of the dark matter phenomenology.
Planck measurements of the power spectrum of the cosmic microwave background radiation infer that the cold dark matter abundance should be around 27\% ($\Omega_{DM} h^2=0.12$), where $h$ is a parameter that accounts for uncertainties in the Hubble rate \cite{Ade:2013zuv}. This alone strongly constrains the viable parameter space of dark matter models.
The observation of cosmic rays and gamma rays also offers a compelling probe for dark matter \cite{Hooper:2012sr,Bringmann:2012ez,Calore:2013yia,Kopp:2013eka,Galli:2013dna,Gomez-Vargas:2013bea,Berlin:2013dva,Madhavacheril:2013cna,
Abazajian:2014fta,Bringmann:2014lpa,Gonzalez-Morales:2014eaa,Buckley:2015doa,Huang:2016pxg}. In particular, the Fermi-LAT sensitivity to continuous gamma-ray emission from dark matter annihilations taking place in Dwarf Galaxies resulted in restrictive bounds in the annihilation cross section today, namely $\sigma v < 3 \times 10^{-26} {\rm cm^3/s}$ for masses of 80 GeV and annihilation into $b\bar{b}$ quark pairs~\cite{Ackermann:2015zua}. This rules out a multitude of light WIMP (weakly interacting massive particles) models in which velocity-independent interactions occur. 

Moreover, underground detectors using liquid XENON, such as XENON \cite{Aprile:2012nq} and LUX \cite{Akerib:2013tjd} that use scintillation and ionization measurements to discriminate signal from background events, observed no excess, leading to the exclusion of spin-independent WIMP-nucleon scattering cross sections larger than $10^{-45} {\rm cm^2}$ for WIMP masses of $50$~GeV. Other experiments have placed complementary limits in particular at lower masses such as SUPERCDMS, which uses Ge targets \cite{Agnese:2015nto}. The ongoing XENON1T \cite{Aprile:2015uzo} and LZ \cite{Malling:2011va} experiments are expected to bring down the limits by roughly two orders of magnitude in the absence of signal and zero background events. 

Besides the indirect and direct detection probes, the Tevatron \cite{Bai:2010hh} and the LHC \cite{Goodman:2010ku,Fox:2011pm} have proven to be great laboratories to test dark matter models. In the case where the dark and visible sectors are connected by vector mediators, dijet \cite{An:2012va,Frandsen:2012rk,Alves:2013tqa,Fairbairn:2016iuf} and dilepton \cite{Profumo:2013sca,Alves:2015pea,Alves:2015mua,Allanach:2015gkd,Kahlhoefer:2015bea} bounds are by far the most stringent constraints. 
Dark matter phenomenology is then dictated by gauge interactions which are determined, once the gauge group behind the origin of the vector mediator is known. The common approach is to consider simplified lagrangians that encompass both Dirac and Majorana dark matter fermions and then to compute dark matter observables; namely, relic density, annihilation and scattering cross sections, the latter being spin-independent and spin-dependent for Dirac and Majorana fermions, respectively\footnote{Dirac fermions also induce spin-dependent interactions but the spin-independent ones lead to stronger constraints.}. The simplified dark matter model approach is interesting, intuitive and serves as a guide for future work. However, they might lead to different results once embedded in a complete theory.

In the context of the B-L model, dark matter scenarios have been previously investigated. In~\cite{Li:2010rb}, the authors discussed the radiative see-saw mechanism to account for neutrino masses and focused exclusively on dark matter abundance. Supersymmetric B-L extensions~\cite{Khalil:2008ps,Burell:2011wh,Basso:2012gz} and a conformal approach~\cite{Okada:2012sg} have also been investigated. Even though later disfavored in~\cite{Mambrini:2015sia}, a global B-L symmetry has been proposed~\cite{Baek:2013fsa}. In~\cite{El-Zant:2013nta} a warm dark matter scenario was investigated. 
The possibility of having one of the right-handed neutrinos to be the dark matter candidate was entertained in \cite{Sahu:2005fe,Basak:2013cga,Kaneta:2016vkq}, whereas in \cite{Rodejohann:2015lca} an additional scalar played this role. This extra scalar dark matter was also investigated in \cite{Guo:2015lxa}, but in the context of classical scale invariance.
The authors of \cite{Sanchez-Vega:2015qva,Patra:2016ofq} considered an exotic B-L model and advocated the presence of many scalar fields. 
Finally, the authors of \cite{Duerr:2015wfa} studied Dirac fermion dark matter in the context of a $U(1)_{B-L}$ symmetry, but with the inclusion of LEP bounds only they discussed gamma-ray lines emissions, which turned out to be irrelevant unless one lives very close to the resonance with a dark matter quantum number under B-L larger than three.

Thus, our work supplements previous studies for the following reasons:

(i) Both fermionic and scalar dark matter realizations are discussed as well as several quantum numbers and gauge couplings options.

(ii) We investigate two-component dark matter scenarios.

(iii) We perform a detailed collider study at next-to-leading order (NLO) plus next-to-leading logarithmic (NLL) accuracy using recent dilepton data from the LHC at 13 TeV, which are often ignored due to the handy LEP limits.

(iv) Finally, the region of parameter space allowed/excluded by limits from the LHC, LEP and indirect detection experiments in dependence of the mass of the mediator, gauge couplings and dark matter mass is presented.

\section{Model}
\label{sec:model}

In the Standard Model, both baryon and lepton numbers are accidental global symmetries.  Thus, a natural extension of the SM consists of gauging both quantum numbers. However, only combinations of B-L are free of triangle anomalies. Interestingly, the gauge anomalies $Tr(U(1)_{B-L} SU(2)_L^2),\ Tr(U(1)_{B-L} U(1)_Y^2)$ and $Tr(U(1)_{B-L}^3)$ vanish with the introduction of three right-handed neutrinos having  charge $(-1)$ under B-L. In addition, this also leads to vanishing gravitational anomalies. Therefore, the gauged B-L symmetry naturally addresses neutrino masses through see-saw mechanisms~\cite{Mohapatra:1979ia,Minkowski:1977sc,Schechter:1980gr,Mohapatra:1980yp,Lazarides:1980nt,Keung:1983uu}. There are several ways to accommodate dark matter without spoiling the anomaly cancellation, namely:
\begin{enumerate}[(i)]

\item{ {\it \bf Dirac Fermion Dark Matter - $Z^{\prime}$ Portal with unbroken B-L}: This model introduces a vector-like Dirac fermion charged under $U(1)_{B-L}$ leaving the B-L symmetry unbroken. Dark matter phenomenology is then governed by the $Z^{\prime}$ portal. The new gauge boson mass is generated through the Stueckelberg mechanism, which leads to the following Lagrangian~\cite{Feldman:2006wb,Feldman:2007wj,Feldman:2011ms,Heeck:2014zfa}:

\begin{eqnarray}
&&\mathcal{L} \supset \bar{\chi}\gamma^{\mu}D_{\mu}\chi -M_{\chi}\bar{\chi}\chi -\frac{1}{4} F^{\prime \mu\nu}F_{\mu\nu}^{\prime}  -\frac{1}{2} M_{Z^{\prime}}Z_{\mu}^{\prime}Z^{\prime \mu}+ g_{BL}\sum_{i=1}^3 (\bar{l}\gamma_{\mu}l+\bar{\nu}_i \gamma_{\mu}\nu_i)Z^{\prime\mu}\nonumber\\ 
&&+ \frac{g_{BL}}{3}\sum_{i=1}^6(\bar{q_i}\gamma_{\mu}q_i)Z^{\prime\mu} \nu + y_{ij}\bar{L_i}\tilde{\phi}\nu_{jR}\, ,
\label{Lagrangian1}
\end{eqnarray}
where $D_{\mu}\chi=(\partial_{\mu} +i g_{BL} n_{\chi}Z_{\mu}^{\prime})\chi$. We denote by $\tilde{\phi}$ the isospin transformation of the Higgs doublet, $\phi=(\phi^+, \phi^0)^T$, defined as $\tilde{\phi}= i\sigma_2 \phi$. The dark matter charge, $n_{\chi}$, should be different from $\pm 1$ to prohibit an additional Yukawa term involving $\chi_R$, that would lead to dark matter decay. Note that $M_{Z^{\prime}}$ is not determined by the B-L symmetry and that the right-handed neutrinos acquire mass through the usual Yukawa term. Consequently, the neutrinos are Dirac fermions with their small masses being obtained via suppressed Yukawa couplings. We emphasize that the dark matter stability is guaranteed by B-L symmetry.
}

\item { {\it \bf Dirac Fermion Dark Matter - $Z^{\prime}$ Portal with broken B-L}:

In this scenario one adds a SM singlet scalar, $S$, carrying charge 2 under the B-L symmetry. Dark matter is realized via a vector-like Dirac fermion $\chi$ as follows:

\begin{eqnarray}
&&\mathcal{L} \supset \bar{\chi}\gamma^{\mu}D_{\mu}\chi -M_{\chi}\bar{\chi}\chi -\frac{1}{4} F^{\prime \mu\nu}F_{\mu\nu}^{\prime}  -\frac{1}{2} M_{Z^{\prime}}Z_{\mu}^{\prime}Z^{\prime \mu}+ g_{BL}\sum_{i=1}^3 (\bar{l}\gamma_{\mu}l+\bar{\nu}_i \gamma_{\mu}\nu_i)Z^{\prime \mu}\nonumber\\ 
&&+ \frac{g_{BL}}{3}\sum_{i=1}^6(\bar{q_i}\gamma_{\mu}q_i)Z^{\prime \mu} + y_{ij}\bar{L_i}\tilde{\phi}\nu_{jR}+ \lambda_S \bar{\nu}_R \nu_R S\, ,
\label{Lagrangian2}
\end{eqnarray}

where $v_{BL}$ is the vev of the singlet scalar $S$ and $M_{Z^{\prime}} = 2 g_{BL} v_{BL}$. This mass term arises after spontaneous symmetry breaking of the B-L symmetry through the scalar $S$. The mass of the new gauge boson is generated through the kinetic term of the scalar.

Interestingly, in this procedure the neutrinos are Majorana particles. The right-handed neutrinos have masses determined by the last term in Eq.\ (\ref{Lagrangian2}), whereas the active neutrinos have their masses generated through the usual see-saw type I mechanism. The dark matter stability in this case is assured by a $Z_2$ symmetry remnant from the B-L spontaneous symmetry breaking.

Another possibility would be to give different charges to the three right-handed neutrinos such as $(5,-4,-4)$, which is still anomaly-free. However, several extra fields are then needed to successfully generate neutrino masses~\cite{Ma:2015mjd}. For other different studies based on the B-L gauge symmetry see \cite{Perez:2009mu,Majee:2010ar,Li:2010rb,Basso:2011hn,Okada:2014nea,Abdelalim:2014cxa,Ovrut:2014rba,Banerjee:2015hoa,Dong:2015yra,
Gardner:2016wov,Hammad:2016trm}.

}

\item { {\it \bf Scalar Dark Matter - $Z^{\prime}$ Portal}:

Scalar dark matter in the context of B-L symmetry is also a plausible alternative to accommodate dark matter, since it requires only two new fields: a singlet scalar $S$, with charge $+2$ under B-L, and a scalar $\phi$, as dark matter, which should be charged under B-L with a quantum number different from multiples of $\pm 2$ for stability purposes~\cite{Rodejohann:2015lca}. Taking this into account, the Lagrangian of this model reads

\begin{eqnarray}
&&\mathcal{L} \supset \mu_S S^{\dagger}S + \frac{\lambda_S}{2} (S^{\dagger}S)^2 + \mu_{\phi}^2 \phi^{\dagger}\phi +  \frac{\lambda_{\phi}^2}{2} (\phi^{\dagger}\phi)^2\nonumber\\
&& +\lambda_1 (\phi^{\dagger}\phi)(H^{\dagger}H)+ \lambda_2 (S^{\dagger}S)(H^{\dagger}H)+\lambda_3 (\phi^{\dagger}\phi)(S^{\dagger}S)  + g_{BL}\sum_{i=1}^3 (\bar{l}\gamma_{\mu}l+\bar{\nu}_i \gamma_{\mu}\nu_i)Z^{\prime \mu}\nonumber\\ 
&&+ \frac{g_{BL}}{3}\sum_{i=1}^6(\bar{q_i}\gamma_{\mu}q_i)Z^{\prime \mu} + y_{ij}\bar{L_i}\tilde{\phi}\nu_{jR}+ \lambda_S \bar{\nu}_R \nu_R S\, .
\label{Lagrangian3}
\end{eqnarray}

The dark matter phenomenology~\cite{Rodejohann:2015lca} is determined by both gauge interactions, $\phi^{\dagger}\phi \rightarrow Z^{\prime} \rightarrow \bar{f}f$, and scalar interactions,  $\phi^{\dagger}\phi \rightarrow h \rightarrow \bar{f}f, SS$. In the first case the dark matter phenomenology is strongly related to the gauge coupling and the $Z^{\prime}$ mass. It is very predictive and connected to collider physics. In the second, the scalar potential couplings control dark matter observables and the strong connection to collider physics is lost, such that we will not discuss it further. For a detailed study see e.g.~\cite{Rodejohann:2015lca}.
}
\end{enumerate}
\begin{figure}[!t]
	\centering{}
	\subfloat[s-channel annihilation process\label{fig:diagram1a}]{
		\includegraphics[scale=0.53]{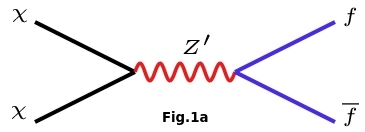}}\hspace{1.5cm}
	\subfloat[t-channel annihilation process\label{fig:diagram1b}]{
		\includegraphics[scale=0.53]{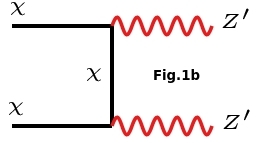}}\hspace{1.5cm}
	\subfloat[Dark matter-nucleon scattering\label{fig:diagram1c}]{
		\includegraphics[scale=0.53]{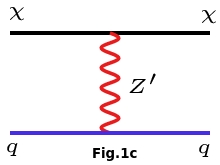}}\hspace{1.5cm}
\caption{Dark matter annihilation and dark matter-nucleon scattering processes in the fermion dark matter model, where $f$ stands for all SM fermions and $q$ represents the quarks. The t-channel annihilation process is only relevant for $M_{\chi} > M_{Z^{\prime}}$.}
\label{fig:diagram1}
\end{figure}

\begin{figure}[!t]
	\centering{}
	\subfloat[s-channel annihilation process\label{fig:diagram2a}]{
		\includegraphics[scale=0.55]{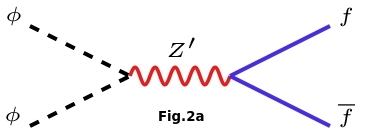}}\hspace{2.5cm}
		\subfloat[Dark matter-nucleon scattering process\label{fig:diagram2b}]{
			\includegraphics[scale=0.57]{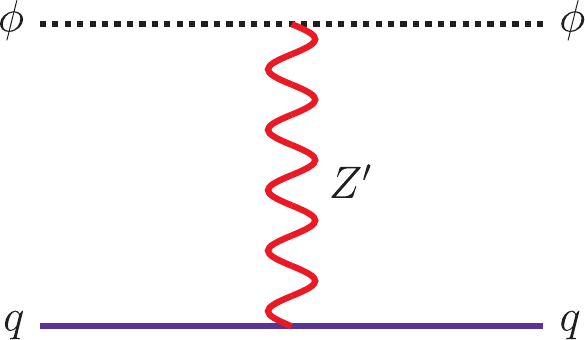}}
\caption{Dark matter annihilation and dark matter-nucleon scattering processes in the scalar dark matter model.}
\label{fig:diagram2}
\end{figure}

\section{Dark Matter Abundance}

The relic abundance of dark matter is determined by solving the Boltzmann equation. The dark matter particle pair annihilates and is pair-produced in equal rate in the early Universe, but as the Universe cools down and expands, eventually the expansion rate approaches the interaction rate, and from then on the dark matter particles are only able to self-annihilate into lighter particles. Eventually, then the expansion rate prevents the dark matter particles from self-annihilating. This episode is referred to as freeze-out. In order words, the abundance of left-over dark matter particles is linked to the annihilation cross section at the freeze-out, which can be very different from the annihilation cross section today ~\cite{Profumo:2013yn}. Thus, the stronger the annihilation cross section is, the fewer remnant dark matter particles subsist today. In what follows, we discuss the abundance of the fermion and scalar dark matter in quantitative terms.

\begin{itemize}

\item {\bf Dirac Fermion}
\end{itemize}

In Figs.~\ref{fig:diagram1a} and \ref{fig:diagram1b}, we show the processes that set the dark matter abundance for the fermion. When $M_{\chi} < M_{Z^{\prime}}$, only the first diagram is relevant. $f$ stands for all SM fermions, including the right-handed neutrinos, whose masses are in the eV range in the case where the B-L symmetry in unbroken, whereas in the broken B-L scenario their masses are kept at $100$~GeV. The precise value for their masses is not relevant, and both cases lead to very similar dark matter phenomenology. For this reason, dark matter observables will be derived without explicitly specifying whether or not the B-L symmetry is broken. 

\begin{figure}[!t]
	\centering{}
	\subfloat[$n=1/3$ and $M_{Z^{\prime}}=4$~TeV\label{fig:abundance1a}]{
		\includegraphics[scale=0.45]{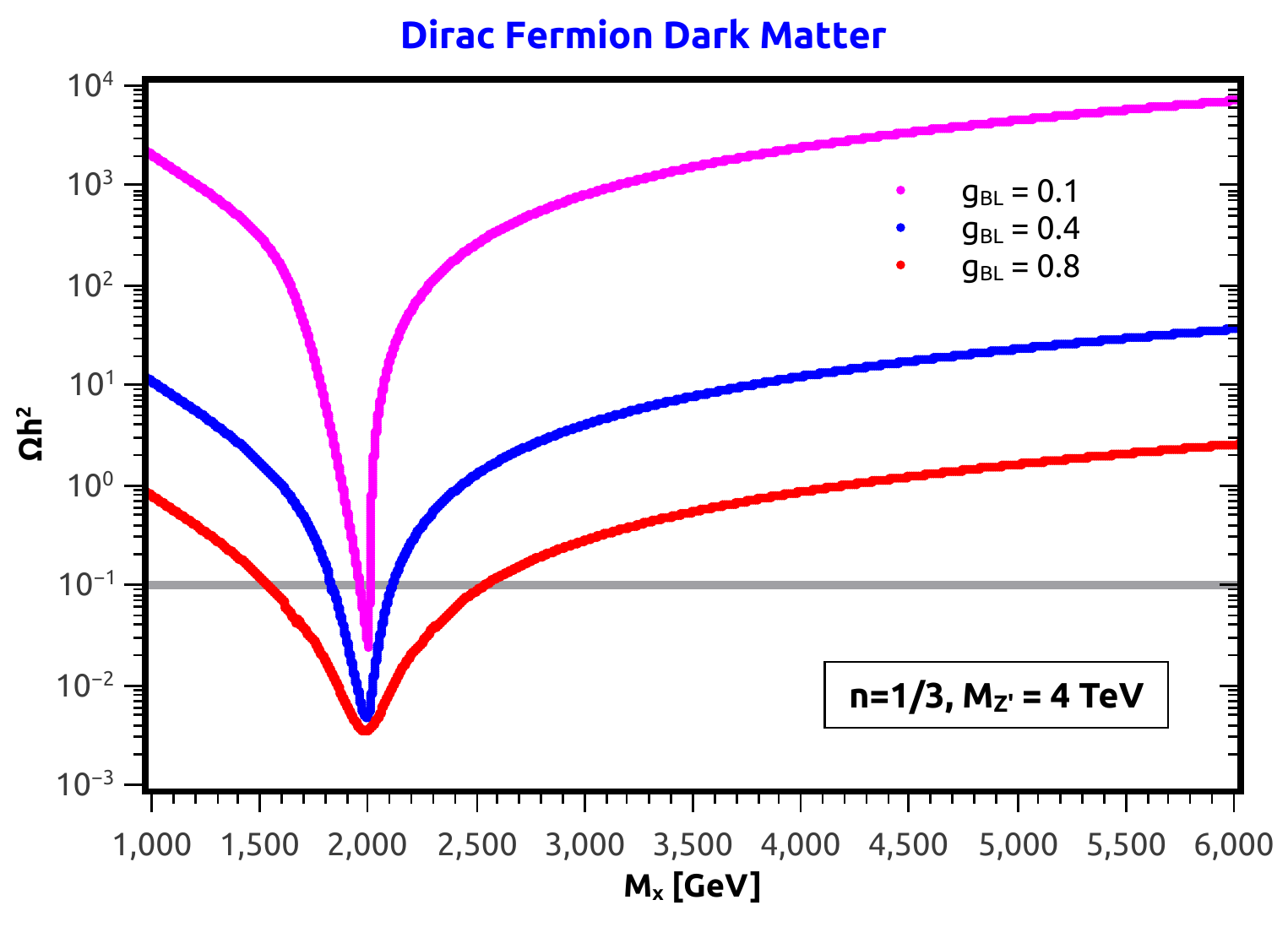}}
		\subfloat[$n=1/3,\ g_{BL}=0.1$\label{fig:abundance1b}]{
			\includegraphics[scale=0.45]{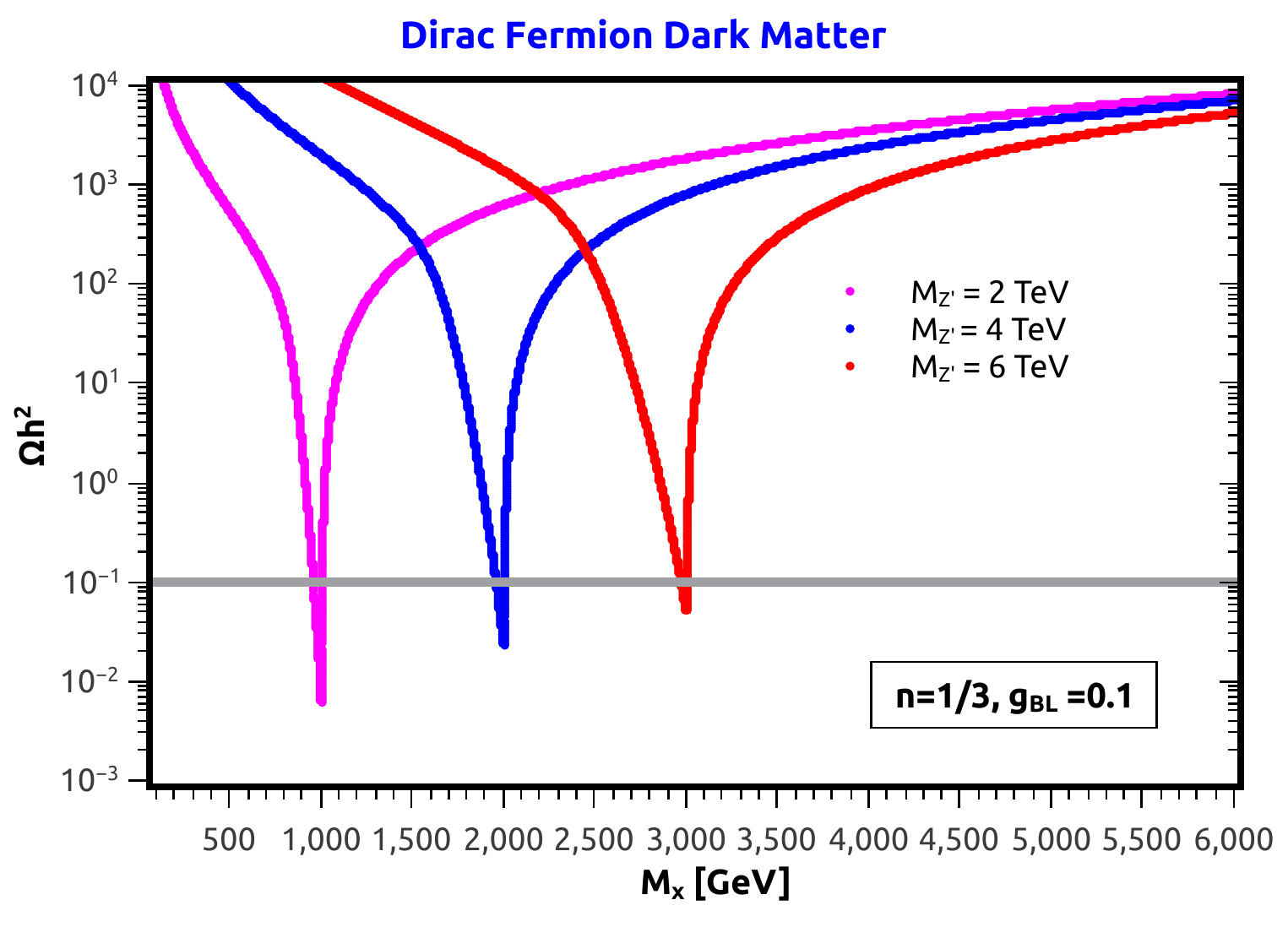}}
\caption{{\it \bf Dirac Fermion}. Abundance as a function of mass for various gauge couplings and $Z'$ boson masses. Because the model must satisfy the relic density, it features a strong dependence in the resonance region.}
\label{fig:abundance1}
\end{figure}

In Fig.~\ref{fig:abundance1} we display, for $n=1/3$, the abundance of the fermion as a function of its mass. In the left panel, Fig.~\ref{fig:abundance1a}, the $Z^{\prime}$ mass has been fixed to $4$~TeV and the gauge coupling varied in $g_{BL}\in[0.1,0.8]$, while in the right panel, Fig.~\ref{fig:abundance1b}, we keep $g_{BL}=0.1$ and vary $M_{Z^{\prime}}=2,4,6$~TeV.

From Fig.~\ref{fig:abundance1a}, it is clear that the increase in the coupling widens the resonance and therefore leads to viable dark matter masses away from $M_{Z^{\prime}}/2$, lower or higher. In addition, the larger the coupling, the larger the annihilation rate, leading to smaller abundance. Thus, one needs sufficiently large gauge couplings to enhance the annihilation rate and reach $\Omega h^2 \sim 0.1$. Notice that the resonance condition is not as needed, if couplings close to unity are used. Such large couplings arise naturally in 3-3-1 models \cite{Mizukoshi:2010ky,Alves:2011mz,Alvares:2012qv,Profumo:2013sca,Kelso:2013nwa,Cogollo:2014jia,
Dong:2014wsa,Dong:2014esa,Kelso:2014qka} and left-right models \cite{Mohapatra:1974gc,Mohapatra:1974hk,Senjanovic:1975rk,Mohapatra:1979ia,
Mohapatra:1980yp,Dias:2010vt,Heeck:2015qra,Rodejohann:2015hka,Deppisch:2015cua,Patra:2015vmp}. Other fermion dark matter models feature similar trends \cite{Esch:2013rta,Esch:2014jpa,Dutra:2015vca,Yaguna:2015mva,Ibarra:2016dlb}. 

The impact of the $Z^{\prime}$ mass is shown in Fig.~\ref{fig:abundance1b}, which exhibits a series of peaks at different dark matter masses. The larger $M_{Z^{\prime}}$ gets, the heavier the dark matter mass has to be in order to achieve the right abundance. We point out that both results for fermion dark matter are presented for $n=1/3$, but they can be easily rescaled, since the abundance scales as $n^2 g_{BL}^4$. Hence, for constant relic density, a change in $n$ straightforwardly induces a quadratically inverse change in $g_{BL}$.

\begin{itemize}

\item {\bf Scalar Field}
\end{itemize}

\begin{figure}[!t]
\centering
\subfloat[$M_{Z^{\prime}}=2$~TeV and $n=1/3$ for $g_{BL}=0.4,0.8,1$.\label{fig:abundance2a}]{
	\includegraphics[scale=0.45]{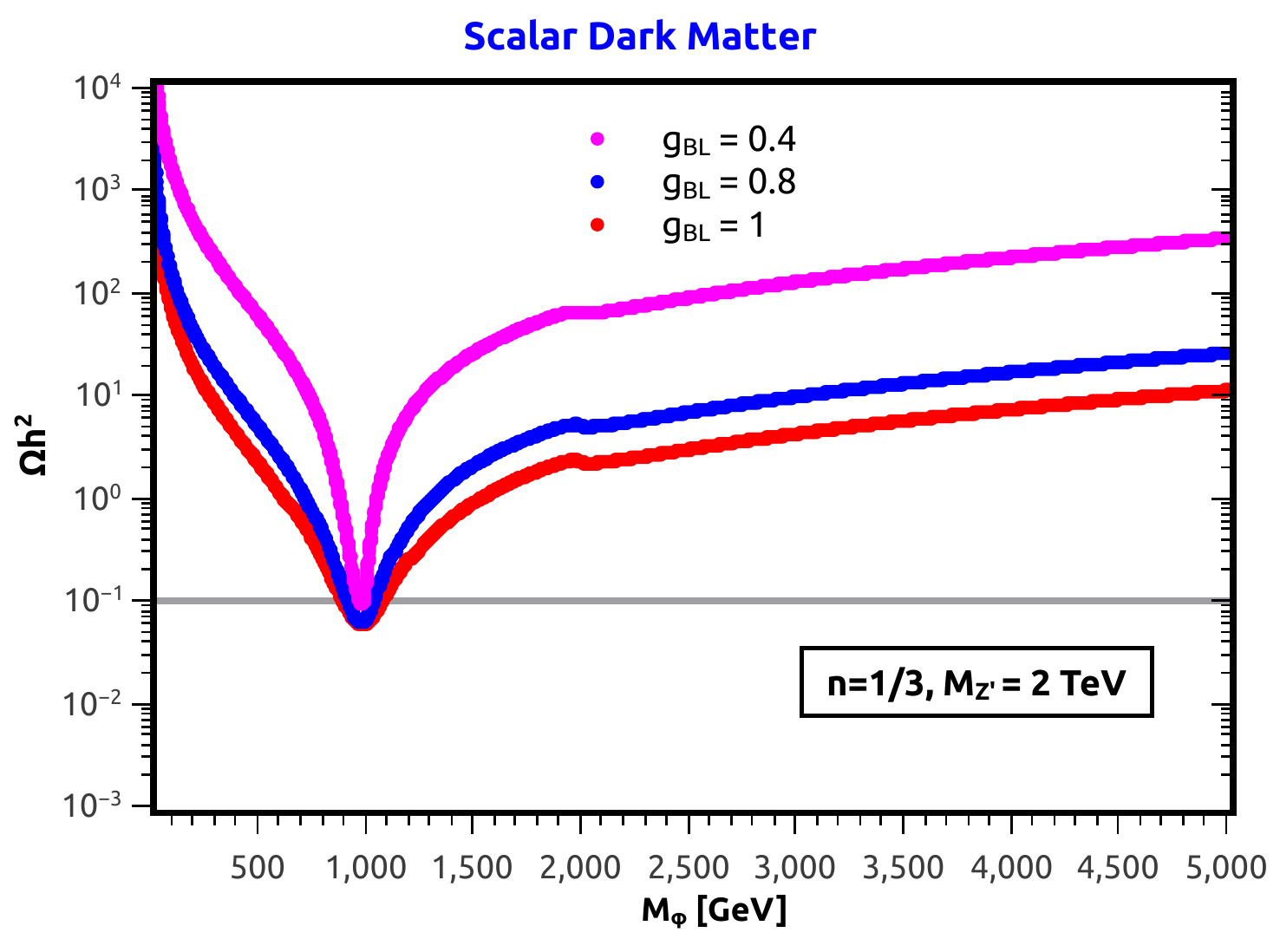}}\hfill
	\subfloat[$M_{Z^{\prime}}=2$TeV and $n=1$ for $g_{BL}=0.4,0.8,1$.\label{fig:abundance2b}]{
		\includegraphics[scale=0.45]{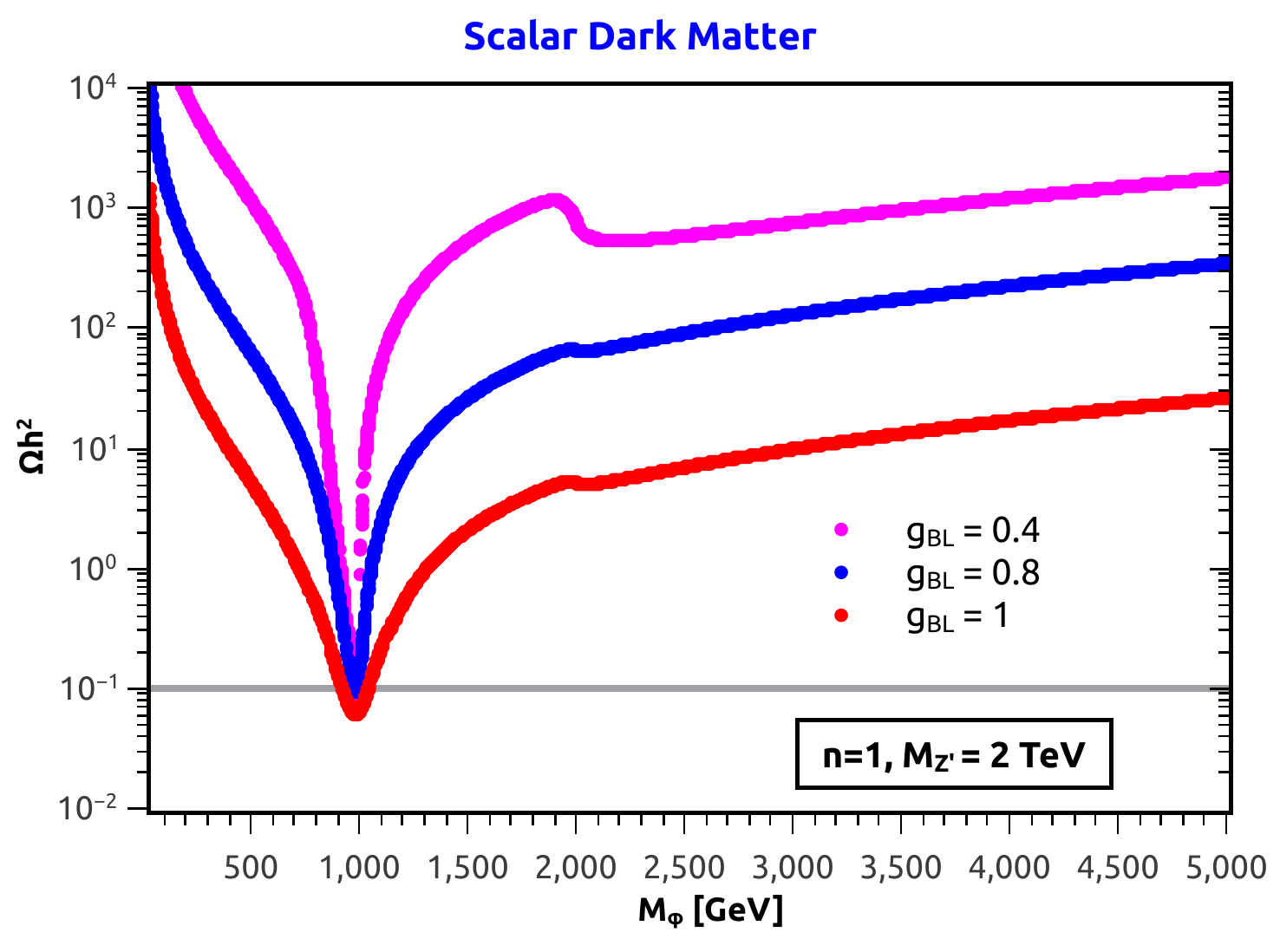}}\\
		\subfloat[$g_{BL}=0.8$ and $n=1$ for $M_{Z^{\prime}}=2,4,6$~TeV.\label{fig:abundance2c}]{
			\includegraphics[scale=0.45]{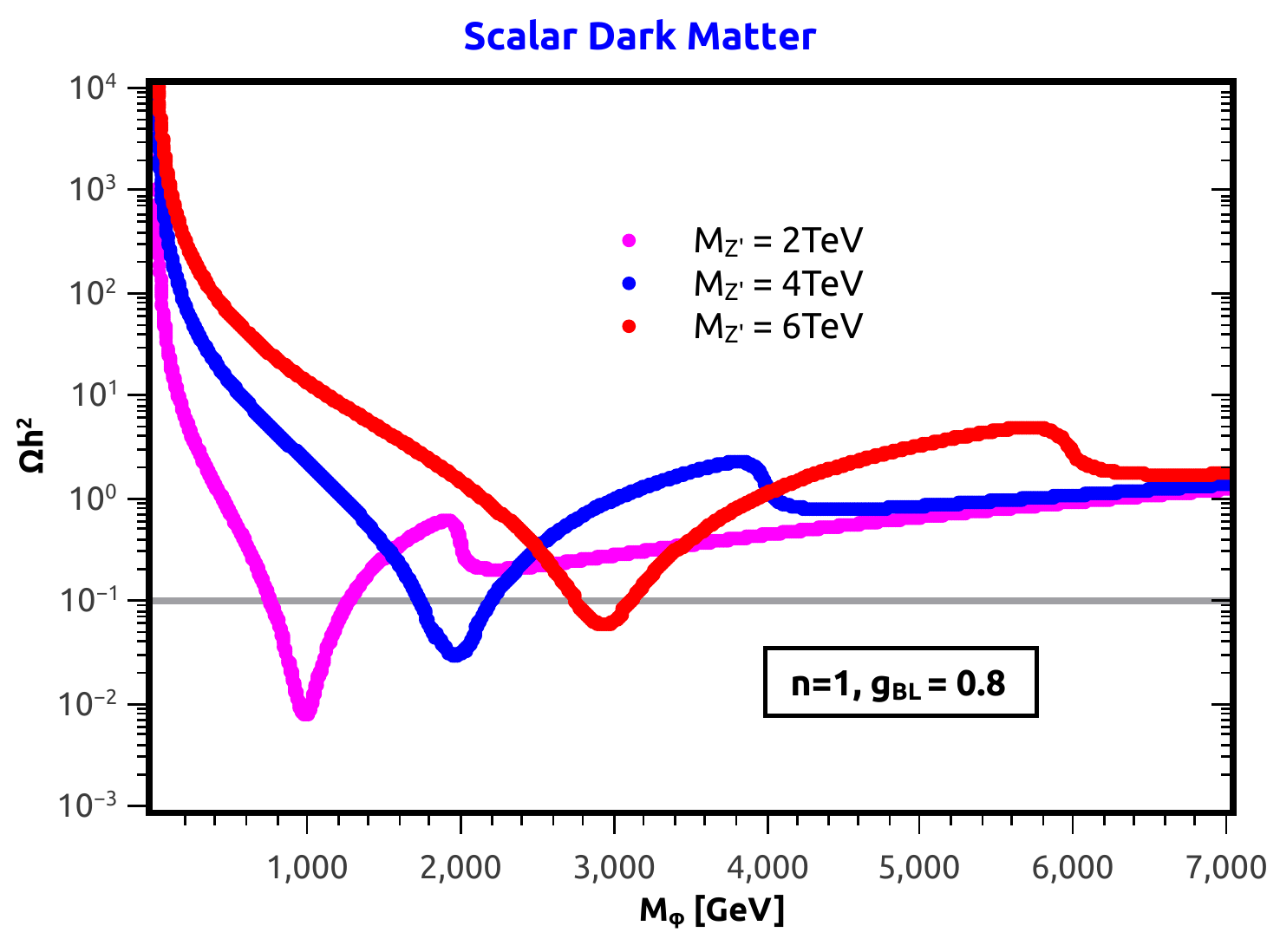}}
\caption{ {\it \bf Scalar Field}. Abundance as a function of mass. The kinks in the plots are the result of the $Z^{\prime}$ threshold, i.e. when the scalar can pair annihilate to produce $Z^{\prime}$ bosons.}
\label{fig:abundance2}
\end{figure}

In Fig.~\ref{fig:diagram2a} we show the Feynman diagram relevant for determining the scalar dark matter abundance. In Fig.~\ref{fig:abundance2} the abundance for two different charges under B-L, $n=1/3$ and $n=1$, is shown. The kinks in the plots are the result of the $Z^{\prime}$ threshold, i.e. when the scalar can pair annihilate into a $Z^{\prime}$ boson\footnote{This effect is also present, but much less pronounced in the fermion case discussed above.}.

Similarly to the case of fermion dark matter, the s-channel resonance regime $m_{\phi} \sim M_{Z^{\prime}}/2$ is responsible for increasing the annihilation cross section and consequently reducing the abundance to values close to the one inferred by Planck. Fig.~\ref{fig:abundance2c} shows the abundance with $n=1$ and $g_{BL}=0.8$ and for various masses of the new gauge boson, $M_{Z^{\prime}}=2,4,6$~TeV. Again, the effect of increasing $M_{Z'}$ is to simply move the resonance region to higher dark matter masses. It is noticeable that for $ g_{BL}=0.8$ the resonance region is wide enough to accommodate two different dark matter masses yielding the right abundance.

As already mentioned, the annihilation cross section grows as $n^2 g_{BL}^4$. 
For $n\ll 1$, one therefore needs gauge couplings larger than one in order to satisfy the relic density constraint. On the other hand, values of $n$ closer to one enhance the dark matter-nucleon scattering rate thus severely restricting the model, as we shall see below.

As a summary, we have seen in this part that both Dirac fermions and scalars can be viable dark matter candidates of the Universe as long as the  annihilation rate occurs not very far from the resonance.

\section{Indirect Dark Matter Detection}

In this B-L model, dark matter self-annihilations take place through vector-like gauge mediation. Therefore, they occur at a similar rate for all SM fermions. That said, one can use gamma-ray observations of dwarf galaxies from the Fermi-LAT satellite to constrain the annihilation cross section into SM fermions, which after hadronization processes produce gamma rays. Fermi-LAT has been able to exclude annihilation cross sections into $b\bar{b}$ of $3\times 10^{-26} {\rm cm^3/s}$ for masses around $1-80$~GeV \cite{Ackermann:2015zua}. There are additional complementary constraints in the literature \cite{Galli:2011rz,Hooper:2012sr,Berlin:2013dva,Gomez-Vargas:2013bea,Kopp:2013eka,Calore:2013yia,Weniger:2013hja,Galli:2013dna,Madhavacheril:2013cna,Gonzalez-Morales:2014eaa,Bringmann:2014lpa,Abazajian:2014fta,Elor:2015bho,Bertoni:2015mla,Buckley:2015doa,Caputo:2016ryl}, which lie in the same ballpark. We therefore decided to adopt the Fermi-LAT collaboration results throughout. In both the fermion and scalar dark matter models, the right relic density is achieved for annihilation cross sections smaller than $3\times 10^{-26} {\rm cm^3/s}$. Since only heavy dark matter particles are viable, much heavier than $100$ GeV, the indirect detection limits are rather subdominant to collider and direct detection ones and for this reason not shown throughout.

\begin{figure}[!t]
\centering
\subfloat[$M_{Z'}=4$ TeV\label{fig:xsectiona}]{
	\includegraphics[scale=0.45]{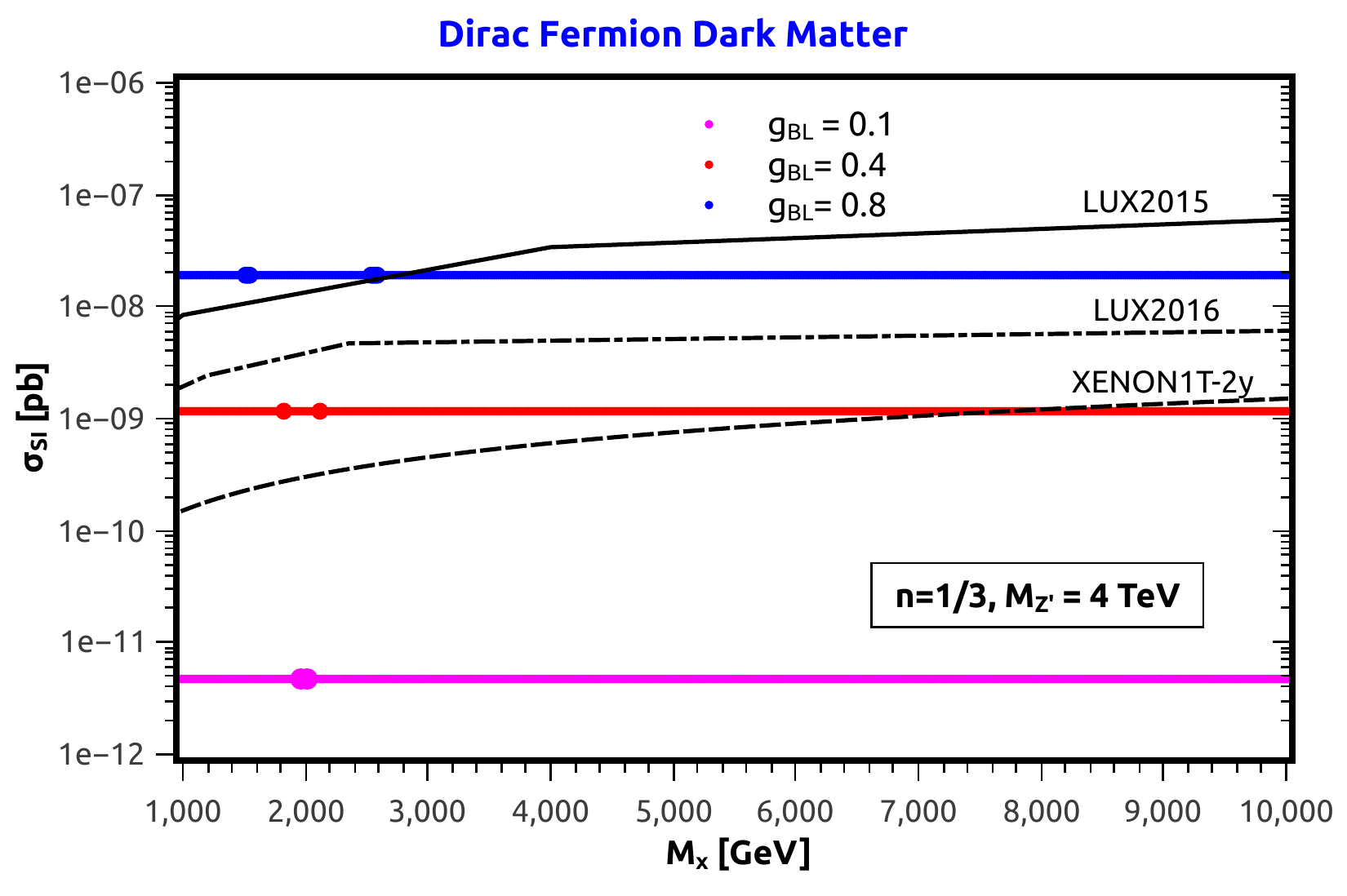}}
	\subfloat[$M_{Z'}=6$ TeV\label{fig:xsectionb}]{
		\includegraphics[scale=0.45]{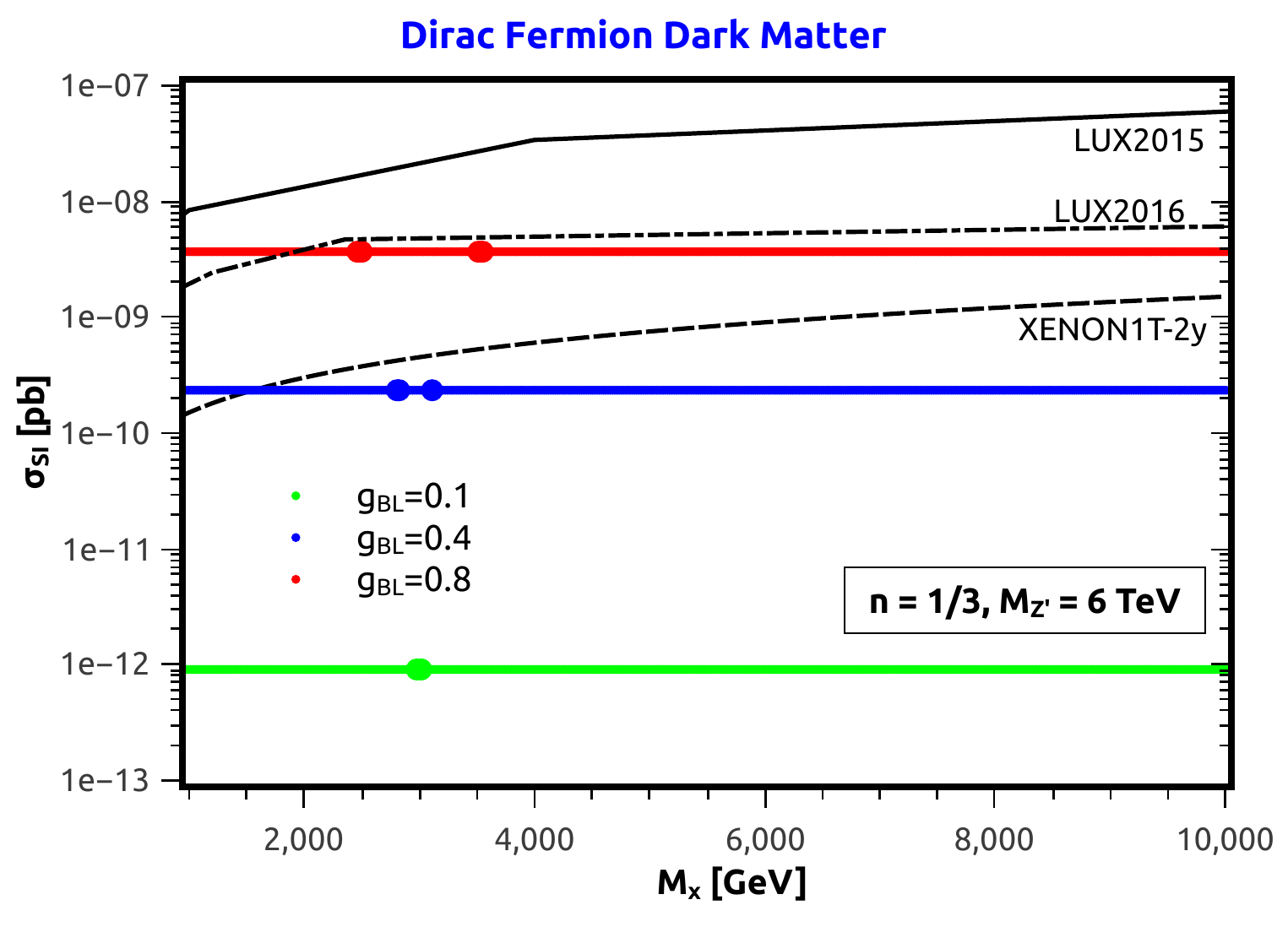}}
\caption{{\it \bf Dirac Fermion}. Spin-independent dark matter-nucleon scattering cross section as a function of the dark matter mass with $n=1/3$ and $M_{Z^{\prime}}=4,6$~TeV for different gauge couplings, $g_{BL}=0.1,0.4,0.8$. The current limit from LUX-2015 (solid line)~\cite{Akerib:2015rjg}, preliminary limit from LUX-2016 (dotted-dashed)\cite{LUX2016} and the one projected from XENON1T (dashed line) for two years of data taking~\cite{Aprile:2015uzo} are also shown.}
\label{fig:xsection}
\end{figure}

\begin{figure}[!t]
\centering
\subfloat[$M_{Z'}=2$ TeV and $g_{BL}\in\{0.1,0.4,0.8\}$\label{fig:xsectionscalara}]{
	\includegraphics[scale=0.45]{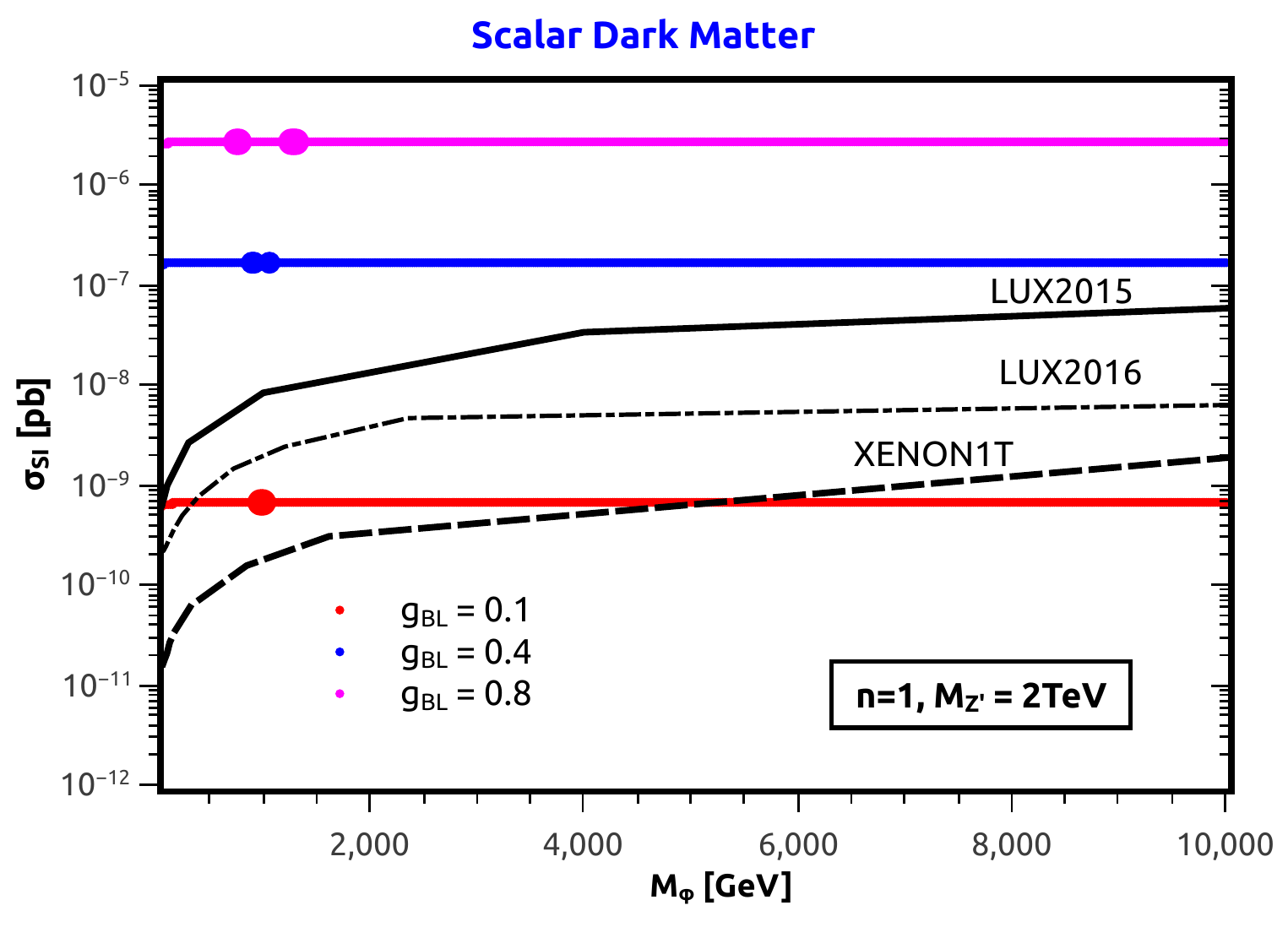}}
	\subfloat[$g_{BL}=0.4$ and $M_{Z'}\in\{2,4,6\}$ TeV\label{fig:xsectionscalarb}]{
	\includegraphics[scale=0.45]{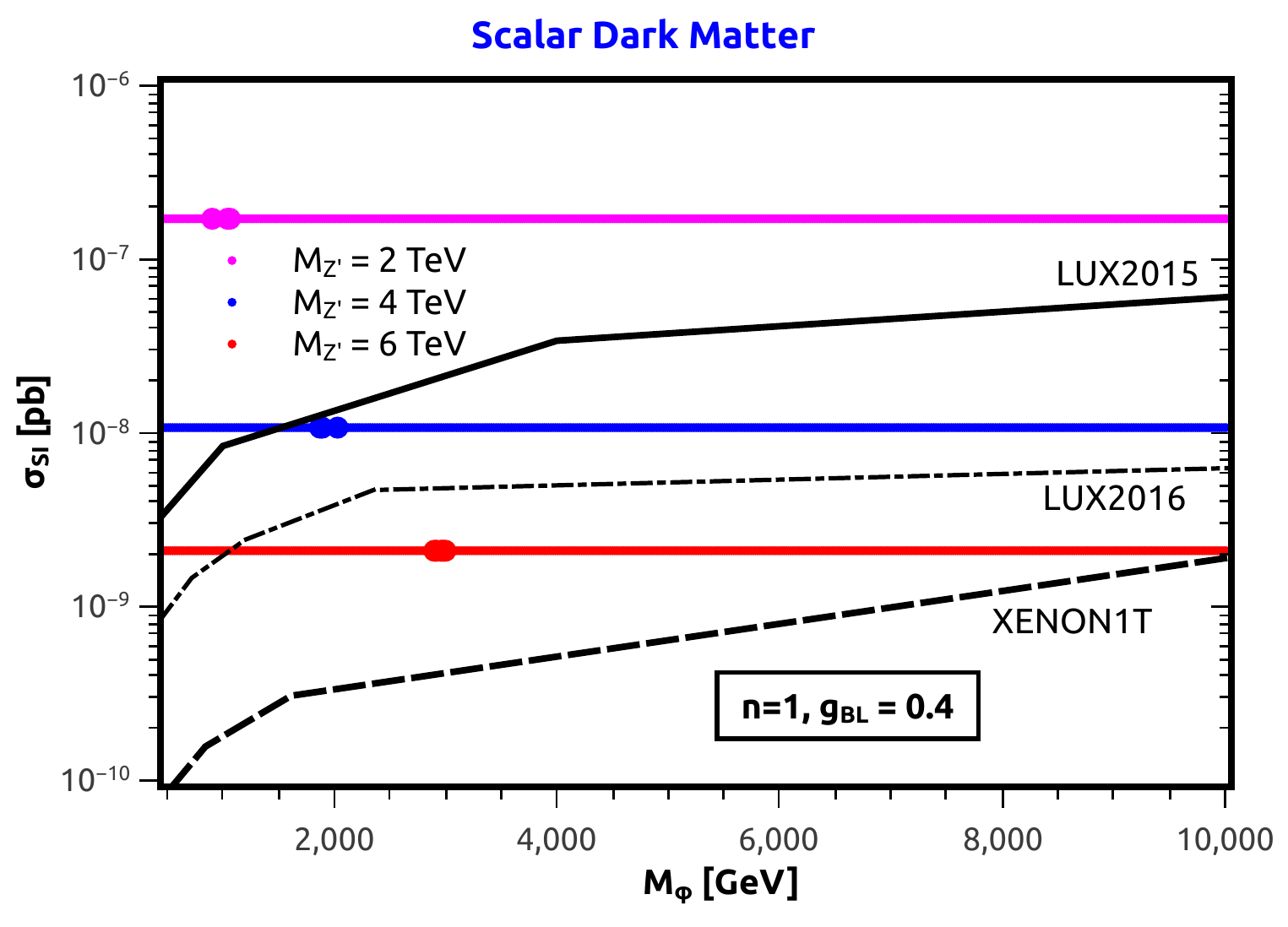}}
	\caption{{\it \bf Scalar Field}. Scattering cross section as a function of the dark matter mass for $n=1$ and various values of $M_{Z'}$ and coupling $g_{BL}$. Predictions are compared to current and projected bounds from LUX-2015 (solid), LUX-2016 (dotted-dashed) and XENON1T (dashed).}
	\label{fig:xsectionscalar}
\end{figure}

\section{Direct Dark Matter Detection}

Direct dark matter detection relies on the measurement of nuclear recoil energies down to energies below 10~keV. The method is based on the use of discriminating variables such as ionization, heat, and scintillation efficiencies to disentangle possible dark matter events from nuclear background rates and mis-identified electron recoils, see~\cite{Freese:2012xd,Cremonesi:2013bma,Klasen:2015uma,Undagoitia:2015gya,Queiroz:2016awc,Mayet:2016zxu} for recent reviews. The measurement of the recoil energy is translated into the plane dark matter-nucleon scattering cross section vs.\ mass, once the dark matter velocity distribution and the local density is set. Since no excess of events has been observed, only limits in this same plane have been derived. LUX experiment provides the world-leading limits on both the spin-independent and spin-dependent scattering cross sections, with the former being more stringent, which we refer as LUX2015 in the figures. However, LUX just presented their new limit with 332 live days, which improves by a factor of four the latest one \cite{LUX2016}. The limit seems to be preliminary, but we have incorporated in the figures with a dotted-dashed line, labelled as LUX2016.

Since in our setup, both Dirac fermion and scalar dark matter models exhibit larger spin-independent rates, we will use the spin-independent bounds. Moreover, we present the projected bounds from the ongoing XENON1T experiment, which is expected to surpass the LUX2015 sensitivity by two orders of magnitude with two years of data taking \cite{Aprile:2015uzo}. In the following, we discuss the results for dark matter-nucleon scattering cross sections for both candidates.

\subsection{Dirac Fermion}

In Fig.~\ref{fig:diagram1c} we show the Feynman diagram responsible for dark matter-nucleon scattering. Fig.~\ref{fig:xsection} shows the spin-independent dark matter-nucleon scattering cross section as a function of the dark matter mass with $n=1/3$ and $M_{Z^{\prime}}=4$~TeV (\ref{fig:xsectiona}) and $M_{Z'}=6~$TeV (\ref{fig:xsectionb}) for different gauge couplings $g_{BL}\in\{0.1,0.4,0.8\}$. In both figures, current limits from LUX2015 (solid line) \cite{Akerib:2015rjg}, preliminary LUX2016 \cite{LUX2016}, and projected limits from XENON1T (dashed line) are superimposed.
The curves read from top to bottom: blue is for $g_{BL}=0.8$, red for $g_{BL}=0.4$, and pink for $g_{BL}=0.1$. The dark blobs in the figure reproduce the right relic abundance. 

From Fig.~\ref{fig:xsectiona}, it is clear that one needs to use gauge couplings smaller than $0.8$ in order to have a viable dark matter candidate with masses below $2$~TeV. If no dark matter signal is seen, the XENON1T experiment is expected to exclude gauge coupling values larger than $0.4$, if the dark matter mass is demanded to be below $8$~TeV. Ramping up the $Z^{\prime}$ mass to $6$ TeV ameliorates the situation, and couplings as low as 0.8 can be allowed in the entire mass range.
This range will, however, be entirely probed by XENON1T, whereas this experiment will only probe dark matter masses below 1.5 TeV for a coupling of 0.4.

\begin{figure}[!t]
	\centering
	\includegraphics[scale=0.6]{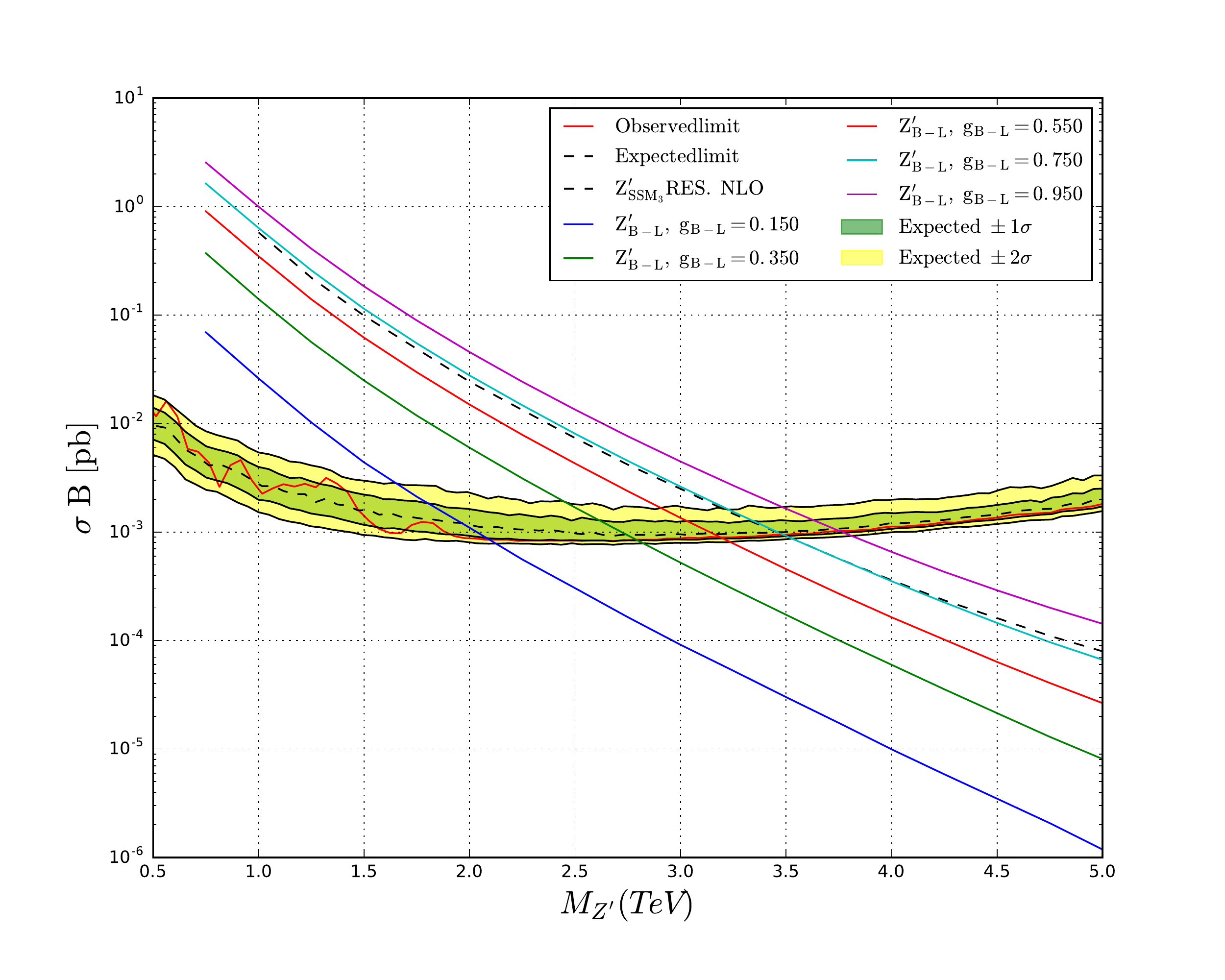}
	\caption{Inclusive total cross section for $pp\rightarrow Z'\rightarrow \ell\bar{\ell}$ at NLO+NLL in the $\mathrm{U(1)}_{B-L}$ models for various values of $g_{B-L}$ as function of the mass of the heavy resonance $M_{Z'}$.}
	\label{fig:atlas_excl}
\end{figure}
\begin{figure}[!h]
	\centering
	\subfloat[Exclusion limit in the plane $M_{Z'}-g_{B-L}$\label{fig:exclusion_contoura}]{
		\includegraphics[width=0.45\textwidth]{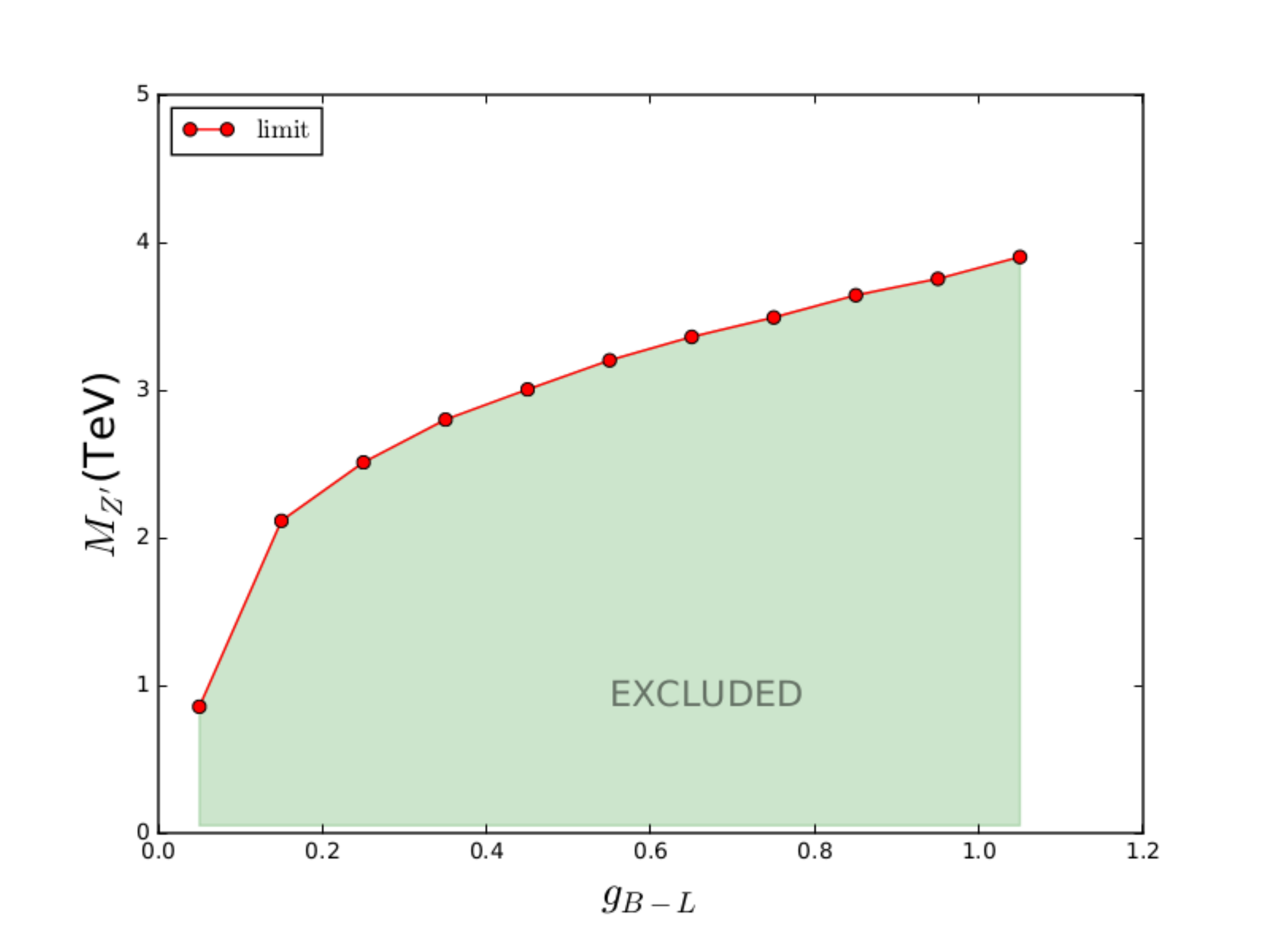}}
	\subfloat[Exclusion limit in the plane $M_{Z'}/g_{B-L}-g_{B-L}$\label{fig:exclusion_contourb}]{
		\includegraphics[width=0.45\textwidth]{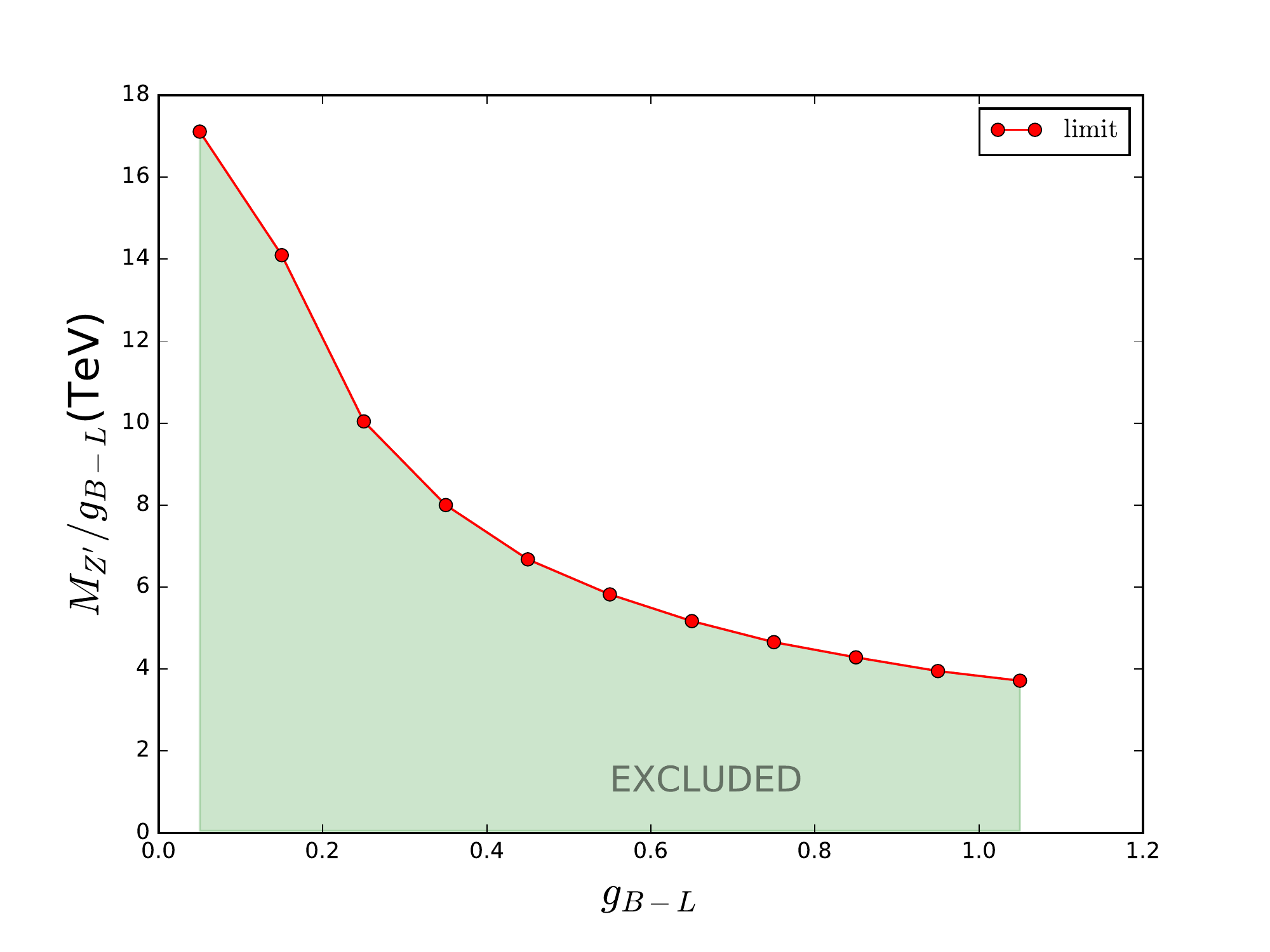}}
		\caption{LHC exclusion limits for the $\mathrm{U(1)}_{B-L}$ model.}
	\label{fig:exclusion_contour}
\end{figure}

\subsection{Scalar Field}

In Fig.~\ref{fig:xsectionscalar} we display the scattering cross section as a function of the dark matter mass with $n=1$ and various values of the new gauge boson mass, $M_{Z^{\prime}}\in\{2,4\}$ TeV, and gauge couplings, $g_{BL}\in\{0.1,0.4,0.8\}$. In both plots, Figs.~\ref{fig:xsectionscalara} and~\ref{fig:xsectionscalarb}, the predictions are compared with current bound from LUX2015 (solid), preliminary from LUX2016 (dotted-dashed), and projected from XENON1T (dashed). The blobs represent points with the right relic density. The value of $n=1$ has been selected in order to simplify the identification of points satisfying the correct dark matter abundance. As before, results can be rescaled taking into account the scaling of the scattering cross section, $n^2 g_{BL}^4/M_{Z^{\prime}}^4$. That is, the result for $n=1,\ g_{BL}=0.4$, is equivalent to the one with $n=1/3$ and $g_{BL}=0.7$.

From Fig.~\ref{fig:xsectionscalara}, one sees that LUX2015 already rules out a large region of the model parameter space, forcing the use of suppressed gauge couplings, e.g. $g_{BL} \sim 0.1$, for $M_{Z^{\prime}}=2$~TeV. Note also that the projected limits from XENON1T might fiercely exclude couplings larger then 0.1. 

Similarly, Fig.~\ref{fig:xsectionscalarb} shows the spin-independent cross section as a function of the dark matter mass for various values of $M_{Z'}$ and fixed $g_{BL}=0.4$ and $n=1$. The LUX experiment excludes $Z^{\prime}$ masses above $4$~TeV, whereas XENON1T has the potential to rule out masses larger than $6$~TeV,  which is in the ballpark of the LHC-14 TeV sensitivity to gauge bosons with an integrated luminosity of  300 fb$^{-1}$~\cite{Ferrari:2000sp,Lindner:2016lxq}. Analogous conclusions would be drawn for $n=1/3$ by simply shifting the gauge coupling as mentioned before.

It is important to keep in mind that collider bounds on the model have been ignored up to now. Including them would lead to the exclusion of some of the points considered above. These limits will be included later on, when we present our results in a more informative plane, that is, $M_{Z^{\prime}}$ vs.\ $g_{BL}$. In what follows, we derive updated limits on the mass of a new neutral gauge boson using $13$~TeV dilepton data from the LHC and compare with the well known LEP bounds.

\begin{figure}[!t]
\centering
\includegraphics[scale=0.8]{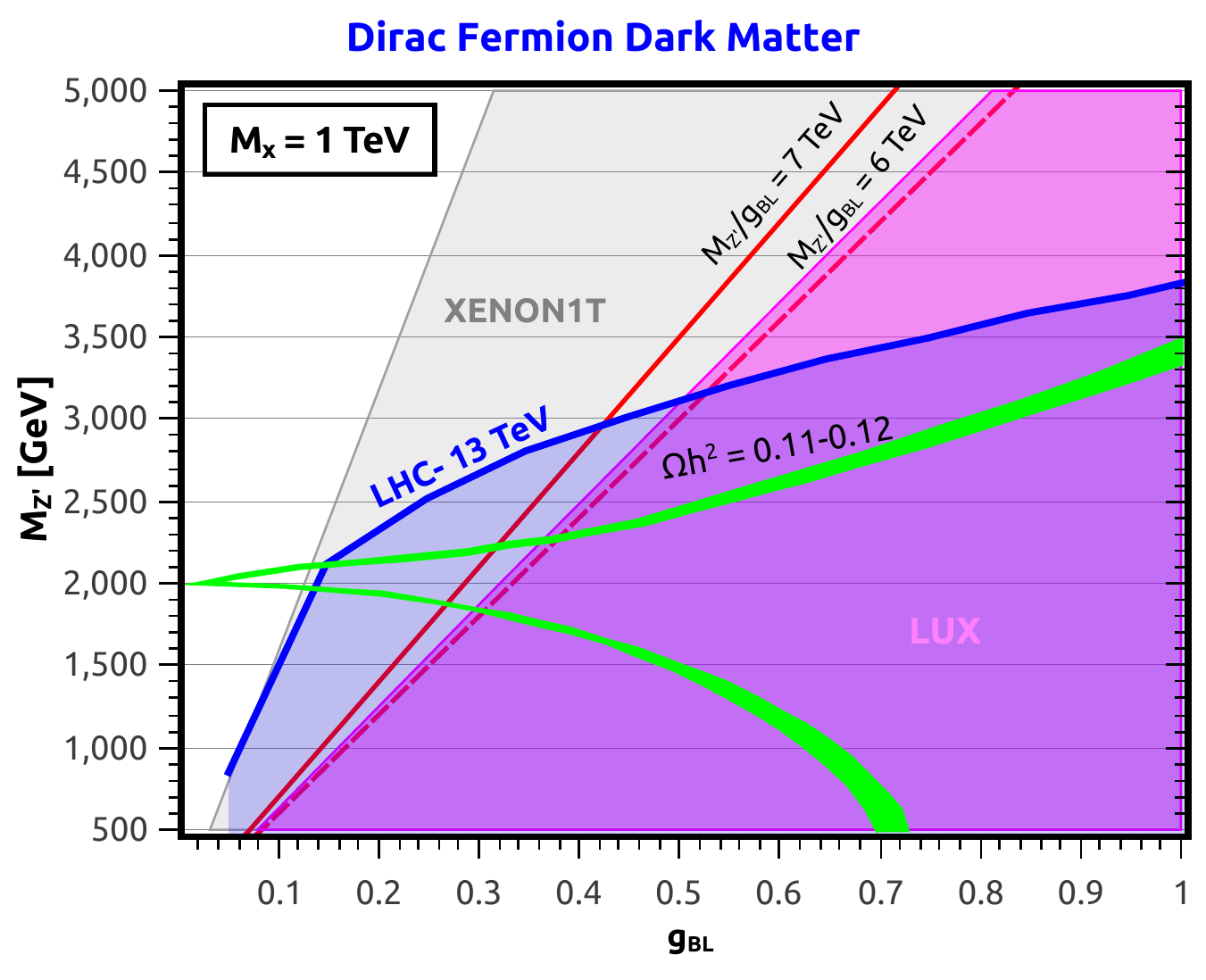}
\caption{Allowed region of parameters for a $1$~TeV Dirac fermion as dark matter. The green curve delimits the region of parameter space with the right abundance ($\Omega h^2=0.11-0.12$), the pink (gray) shaded region is ruled out by LUX2015 (XENON1T), the blue region is excluded by dilepton data from the LHC, and the solid red (dashed) lines represent the current (old) LEP-II bounds, namely $M_{Z^{\prime}}/g_{BL}> 7$~TeV ($M_{Z^{\prime}}/g_{BL}> 6$~TeV).}
\label{fig:5}
\end{figure}

\begin{figure}[!t]
\centering
\includegraphics[scale=0.8]{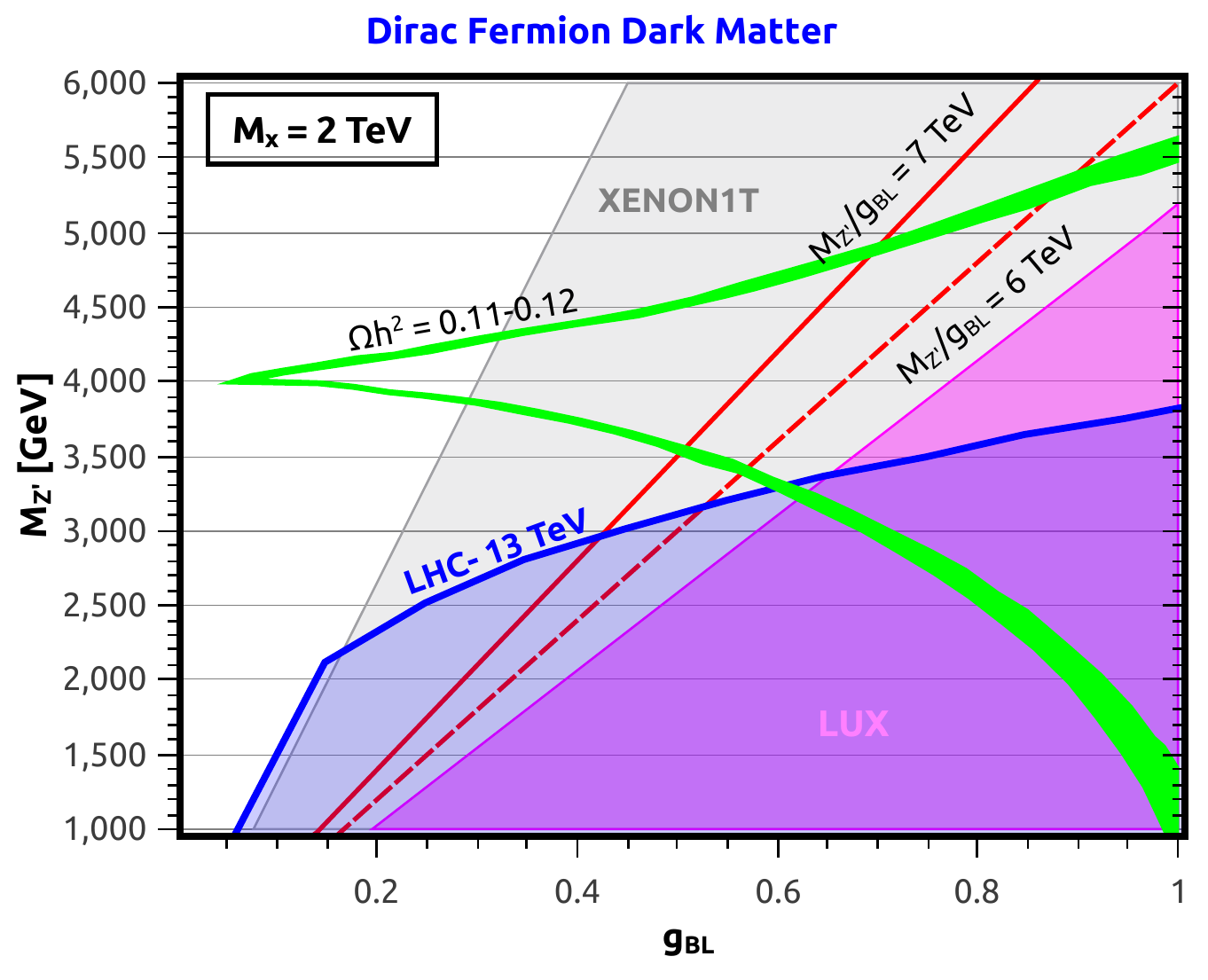}
\caption{Allowed region of parameters for a $2$~TeV Dirac fermion as dark matter. The green curve delimits the region of parameter space with the right abundance ($\Omega h^2=0.11-0.12$), the pink (gray) shaded region is ruled out by LUX2015 (XENON1T), the blue region is excluded by dilepton data from the LHC, and the solid red (dashed) lines represent the current (old) LEP-II bounds, namely $M_{Z^{\prime}}/g_{BL}> 7$~TeV ($M_{Z^{\prime}}/g_{BL}> 6$~TeV).}
\label{fig:6}
\end{figure}

\begin{figure}[!t]
\centering
\includegraphics[scale=0.8]{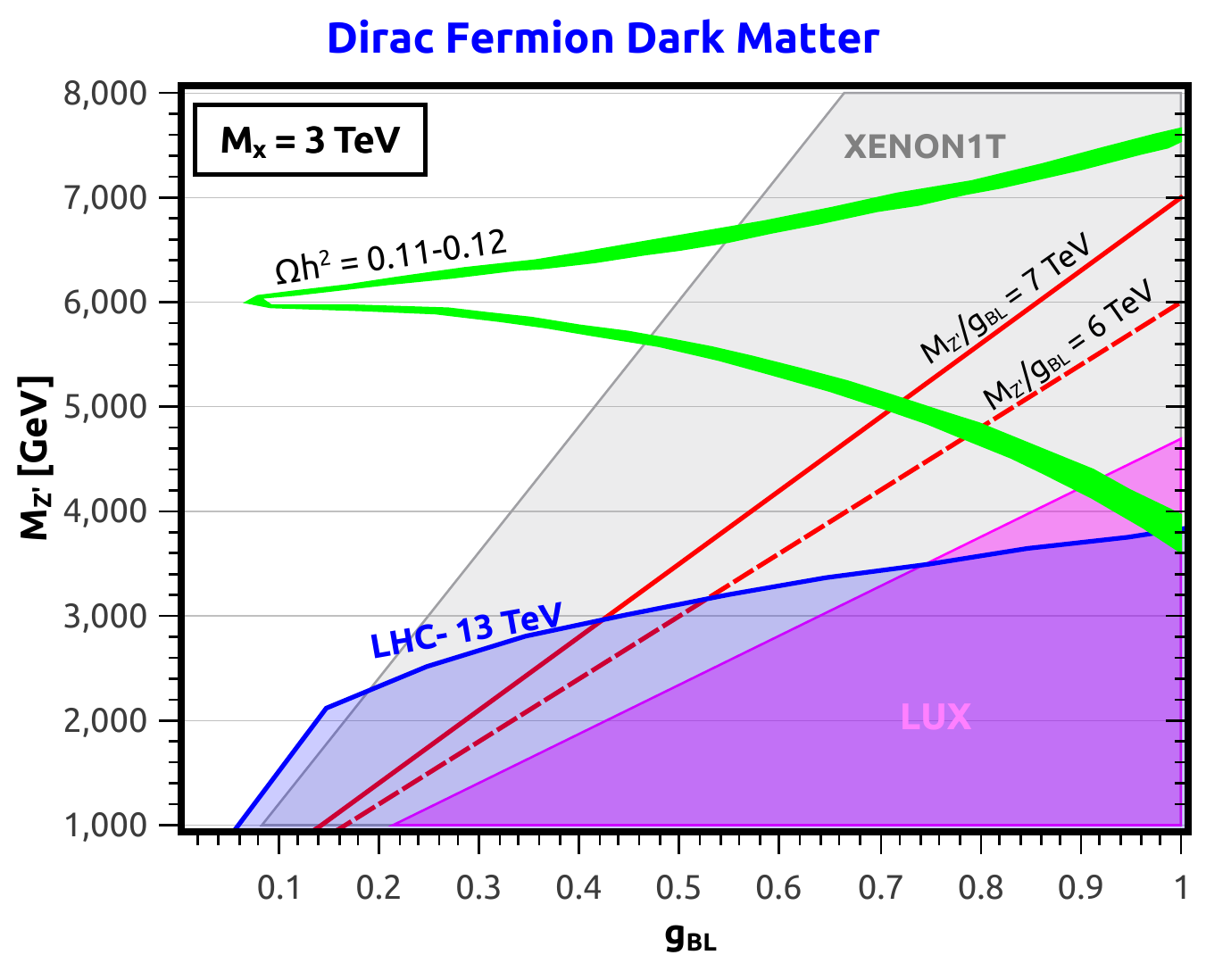}
\caption{Allowed region of parameters for a $3$~TeV Dirac fermion  as dark matter. The green curve delimits the region of parameter space with the right abundance ($\Omega h^2=0.11-0.12$), the pink (gray) shaded region is ruled out by LUX2015 (XENON1T), the blue region is excluded by dilepton data from the LHC, and the solid red (dashed) lines represent the current (old) LEP-II bounds, namely $M_{Z^{\prime}}/g_{BL}> 7$~TeV ($M_{Z^{\prime}}/g_{BL}> 6$~TeV).}
\label{fig:7}
\end{figure}

\section{Collider Limits}
\label{sec:coll_limits}

The ATLAS and CMS collaborations have performed extensive analyses to search for new heavy resonances in both dilepton and dijet signals. In the absence of any excess event over the Standard Model background, the two experiments derived lower bounds on the mass of the \PZP-boson, with dileptons offering stronger limits than dijets due to relatively fewer background events. These bounds are limited to a given model, and typically the experiments express their results assuming simplified models such as the Sequential SM (SSM) or the GUT-inspired $\mathrm{E}_6$ models. 

In this work, however, we re-interpreted their results in terms of the B-L model in question\footnote{See also \cite{Batell:2016zod} for displaced vertices limits in the B-L model, which are weaker for the region of interest.}. In particular, the ATLAS collaboration~\cite{ATLAS-CONF-2015-070} analyzed 3.2 fb$^{-1}$ of $pp$ collisions at $\sqrt{s}=13$ TeV searching for new phenomena in the dilepton final state and extracted the limit $M^{SSM}_{Z'}\geq 3.4$ TeV\footnote{Note that the width of the heavy resonance was fixed to $3\%$ of its mass.}. To calculate the total production cross section of a heavy neutral resonance \PZP and its subsequent decay into leptons, we use the public code RESUMMINO~\cite{Fuks:2013vua}, in which we implemented the appropriate couplings. RESUMMINO implements threshold resummation for total cross sections, $p_T$-resummation for the $p_T$ distribution of heavy gauge bosons, as well as a joint resummation matched to the fixed-order NLO calculation.

When it comes down to interpreting dilepton resonance searches from ATLAS to a model different from the ones aforementioned, one needs to carefully compute the propagator width. In the B-L model, the width, $\Gamma_{Z^{\prime}}$ is proportional to $g_{BL}^{2} M_{Z'}$ and was estimated using PYTHIA 8.215~\cite{Sjostrand:2006za, Sjostrand:2007gs}. It was found to follow the relation 
\begin{equation}
	\frac{\Gamma_{Z'}(g_{BL})}{M_{Z'}}= \frac{\Gamma_{Z'} (g_{BL}=0.7)}{M_{Z'}}\left(\frac{g_{BL}}{0.7}\right)^{2}= 3\%\left(\frac{g_{BL}}{0.7}\right)^{2}\,
	\label{eq:gamma_rel}
\end{equation}
to a very good precision. Therefore, it is clear that for any perturbative values of $g_{BL}$, the $Z'$-boson can be considered as a narrow resonance. For our numerical study we use the CT14~\cite{Dulat:2015mca} NLO PDF set with $\alpha_{S}(M_Z)=0.118$. Following~\cite{ATLAS-CONF-2015-070}, we cut on the transverse mass of the lepton pair, $q_{\ell\ell}^2 \geq 500$ GeV. For each value of mass, $M_{Z'}$, the electroweak coupling constant $\alpha_{EW}$ is evolved to $\alpha_{EW}(M_{Z'}^2)$. Finally, we set the factorization and renormalization scales such that $\mu_F=\mu_R=M_{Z'}$. With these settings, we were able to reproduce to a good level ($\sim$2-3\%) the ATLAS predictions for the SSM.

In Fig.~\ref{fig:atlas_excl}, we show the inclusive total cross section for the process, $pp\rightarrow Z'\rightarrow \ell\bar{\ell}$ calculated at NLO+NLL for the B-L model for various values of the gauge coupling $g_{B-L}$ and as a function of the mass of the heavy resonance. From this, it is straightforward to estimate the lower bound on the mass of the resonance. In Fig.~\ref{fig:exclusion_contoura}, we exhibit this limit in the plane $M_{Z'}$ vs.\ $g_{BL}$, while Fig.~\ref{fig:exclusion_contourb} shows the same limit in the plane $M_{Z'}/g_{BL}$ vs.\ $g_{BL}$. 

Comparing with the SSM result obtained by ATLAS, we see that the exclusion bound for the B-L model is weaker. Note that in a recent analysis~\cite{Guo:2015lxa} the LHC bounds for $\sim5$ fb$^{-1}$ of data and $8$~TeV center-of-mass energy were computed. The conclusion was that for $M_{Z^{\prime}} < 3$~TeV the LHC bounds are stronger than those from LEP, which is in very good agreement with our results obtained at $13$~TeV with $3.2\, {\rm fb^{-1}}$ of data. For the SSM, ATLAS results for 13 TeV with 3.2 fb$^{-1}$ are a bit stronger than those at $8$~TeV and $20\, {\rm fb^{-1}}$, which uses much more data than the analysis in~\cite{Guo:2015lxa}. In addition to that, our results rely on the inclusion of NLO+NLL order effects, which improves our limits.  Thus, the collider limits in \cite{Guo:2015lxa} seem to be overoptimistic. Moreover, an assessment of the LHC sensitivity to the B-L model at 13 TeV, was recently performed in \cite{Okada:2016gsh} without inclusion of detector effects and NLO corrections. There, the authors have found a limit much stronger than ours, namely $M_{Z^{\prime}} > 3$~TeV for $g_{BL}=0.01$. 

We are now ready to combine the relic density, direct detection and collider constraints in the model. To do so, perhaps it is more informative to gather the results in the plane  $M_{Z^{\prime}}$ vs.\ $g_{BL}$, since these two parameters basically define the B-L symmetry.

\begin{figure}[!t]
\centering
\includegraphics[scale=0.8]{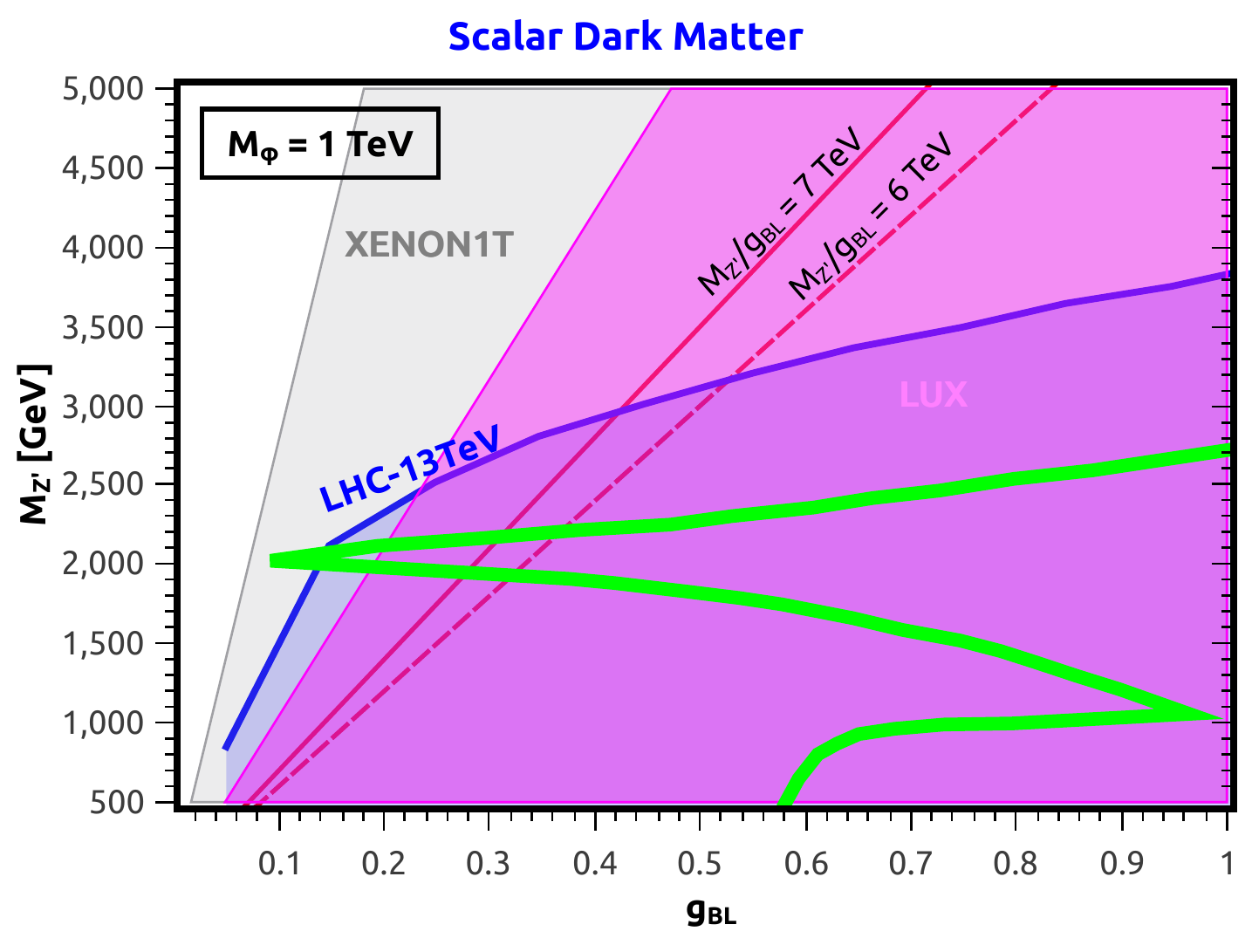}
\caption{Allowed region of parameters for a $1$~TeV scalar field  as dark matter. The green curve delimits the region of parameter space with the right abundance ($\Omega h^2=0.11-0.12$), the pink (gray) shaded region is ruled out by LUX2015 (XENON1T), the blue region is excluded by dilepton data from the LHC, and the solid red (dashed) lines represent the current (old) LEP-II bounds, namely $M_{Z^{\prime}}/g_{BL}> 7$~TeV ($M_{Z^{\prime}}/g_{BL}> 6$~TeV).}
\label{fig:8}
\end{figure}

\begin{figure}[!t]
\centering
\includegraphics[scale=0.8]{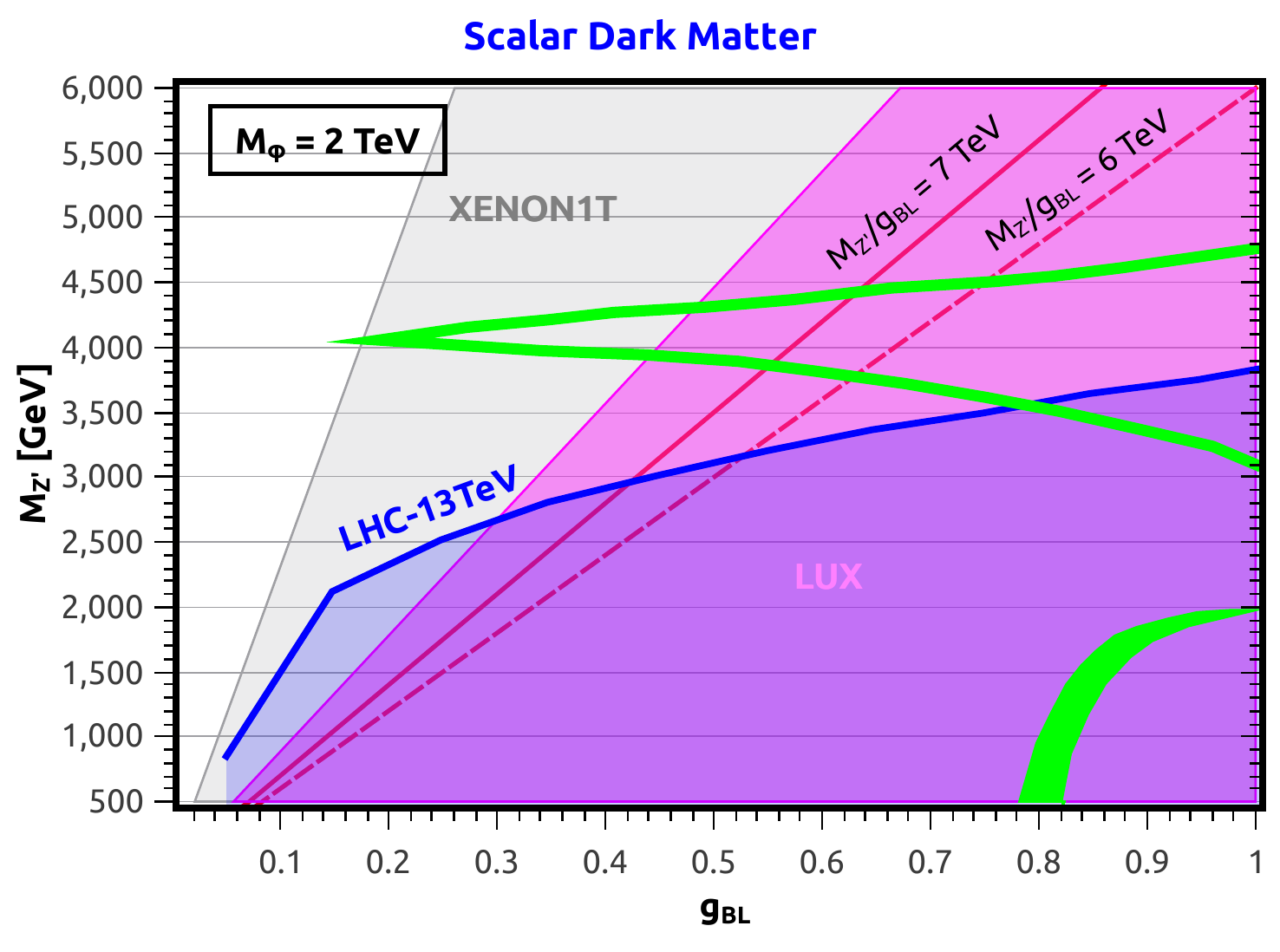}
\caption{Allowed region of parameters for a $2$~TeV scalar field  as dark matter. The green curve delimits the region of parameter space with the right abundance ($\Omega h^2=0.11-0.12$), the pink (gray) shaded region is ruled out by LUX2015 (XENON1T), the blue region is excluded by dilepton data from the LHC, and the solid red (dashed) lines represent the current (old) LEP-II bounds, namely $M_{Z^{\prime}}/g_{BL}> 7$~TeV ($M_{Z^{\prime}}/g_{BL}> 6$~TeV).}
\label{fig:9}
\end{figure}

\begin{figure}[!t]
\centering
\includegraphics[scale=0.8]{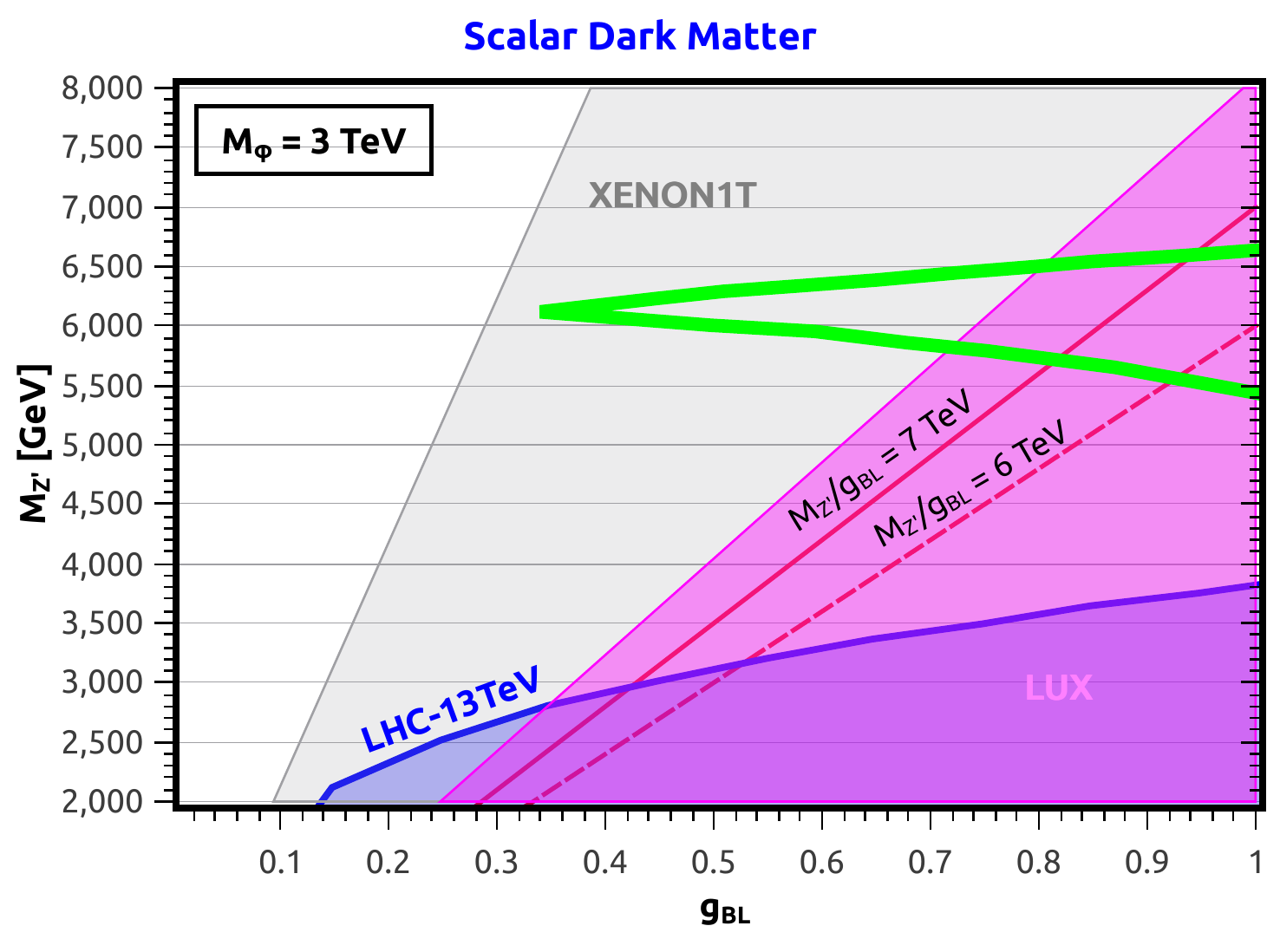}
\caption{Allowed region of parameters for a $3$~TeV scalar field  as dark matter. The green curve delimits the region of parameter space with the right abundance ($\Omega h^2=0.11-0.12$), the pink (gray) shaded region is ruled out by LUX2015 (XENON1T), the blue region is excluded by dilepton data from the LHC, and the solid red (dashed) lines represent the current (old) LEP-II bounds, namely $M_{Z^{\prime}}/g_{BL}> 7$~TeV ($M_{Z^{\prime}}/g_{BL}> 6$~TeV).}
\label{fig:10}
\end{figure}

\section{Combined Results}

\subsection{Dirac Fermion}
In this section we outline the viable parameter space in an arguably more informative plane, i.e. $M_{Z^{\prime}}$ vs.\ $g_{BL}$ with charge  $n=1/3$ under B-L throughout. We combine our findings from relic density, direct detection and collider searches for both the Dirac fermion and scalar dark matter models.

In all figures, the green curve delimits the region of parameter space yielding the right abundance ($\Omega h^2=0.11-0.12$), the pink (gray) shaded region is excluded by LUX2015 (XENON1T), the blue region is ruled out by dilepton data from the LHC, and the solid red (dashed) lines represent the current (old) LEP-II bounds, namely $M_{Z^{\prime}}/g_{BL}> 7$~TeV ($M_{Z^{\prime}}/g_{BL}> 6$~TeV).

In Fig.~\ref{fig:5} we collect these results for a $1$~TeV Dirac fermion, which features a $Z^{\prime}$ resonance of $2$~TeV. Since the annihilation cross section grows with $n^2 g_{BL}^4/(4m_{\chi}^2-M_{Z^{\prime}}^2)^2$, we can see that for small gauge couplings one needs to live very close to the resonance to obtain the right relic density, but as we increase the coupling, the regions relatively far from the resonance become viable. The annihilation cross section is typically small, leading to overabundant dark matter. Therefore one needs to either use large gauge couplings or be near the resonance region to increase the annihilation cross section and bring down the relic abundance to the correct value. Interestingly, LUX2015 limits on the spin-independent scattering cross section exclude a large region of parameter space, especially large values of the coupling. The linear behavior of direct detection limits occurs simply because the scattering cross section scales as $n^2g_{BL}^4/M_{Z^{\prime}}^4$. Consequently larger couplings are more strongly constrained by direct detection, but since $g_{BL}$ and $M_{Z^{\prime}}$ decrease simultaneously in the plane the direct detection limits are simply lines. The inclination is determined by the magnitude of the limit. For instance, XENON1T in two years of data is expected to improve LUX2015 bound by about two orders of magnitude, thus the steeper inclination. It is quite remarkable that XENON1T by itself may rule out almost the entire parameter space of the model. LHC-13 TeV limits based on dilepton data already now exceed the revised LEP-II bound and the LUX sensitivity for this model for gauge couplings smaller than 0.4. 

In Figs.~\ref{fig:6}-\ref{fig:7} similar results for $m_{\chi}=2,3$~TeV are also shown. The model is less constrained as the dark matter mass increases for two reasons: \begin{inparaenum}[(i)]\item the direct detection limits are weakened as a result of fewer dark matter events. Indeed, since the local density, $\rho_{\odot}=n_{\chi}M_{\chi}$, is fixed, we have less dark matter events as we increase the mass; \item the resonance is located at $M_{\chi} \sim M_{Z^{\prime}}/2$ and therefore moves upwards along the $M_{Z^{\prime}}$ axis, towards a weakened LUX and XENON1T limit.\end{inparaenum}

\subsection{Scalar Field}

The possibility of having a singlet scalar dark matter in the B-L model is very much constrained\footnote{As aforementioned, we keep the same color scheme for all figures.}. In Fig.~\ref{fig:8} we present the result for $M_{\phi}=1$~TeV. First, we note that as in the Dirac fermion case, for sufficiently large values of the gauge coupling, there are regions of parameter space away from the $Z^{\prime}$ resonance at $2$~TeV where the correct relic density is achieved. Then, it is clear that there exists a strong degree of complementarity among dilepton, LUX2015 and LEP limits. Combined they fiercely exclude almost the entire parameter space of the model for $M_{\phi}=1$~TeV. 
Only at the resonance is the model capable of satisfying all constraints and reproduce the right dark matter abundance. Strikingly, XENON1T is expected to rule out the possibility of having a $1$~TeV scalar dark matter particle in the B-L model. Note that decreasing the dark matter mass will not be sufficient as the direct detection constraints then get stronger.
Similarly, increasing the scalar mass to around 2-3~TeV does not have much impact as shown in Figs.~\ref{fig:9}-\ref{fig:10}. 
Finally for a mass of 2 TeV, there is a tiny region right at the peak of the $Z^{\prime}$ resonance that might survive the projected XENON1T bound. At this point, the result must be taken with a grain of salt, since the precise XENON1T sensitivity would be required to draw any definite conclusion. Our findings agree approximately with~\cite{Rodejohann:2015lca}, but there the authors used an outdated XENON1T reach.

\subsection{Mixed Dark Matter Scenario}

Two-component dark matter is a plausible scenario. There is no fundamental reason to have one WIMP comprising the entire dark matter of the Universe. In the situation where solid signals come from direct detection and indirect dark matter searches, two-component dark matter arises as a promising framework. Several publications in the past have focused on two- or multi-component dark matter~\cite{Dienes:2011ja,Daikoku:2011mq,Dienes:2012cf,Bhattacharya:2013hva,Kajiyama:2013rla,Biswas:2013nn,Geng:2013nda,
Anandakrishnan:2013tqa,Geng:2014dea,Dienes:2014via,Queiroz:2014ara,Allahverdi:2014ppa,Allahverdi:2014bva,Allahverdi:2014eca,
Biswas:2015sva,Bian:2014cja,Esch:2014jpa,Bae:2014efa,Bae:2015efa,Bae:2015rra,Alves:2016bib}.

\begin{figure}[!h]
\centering
\includegraphics[scale=0.7]{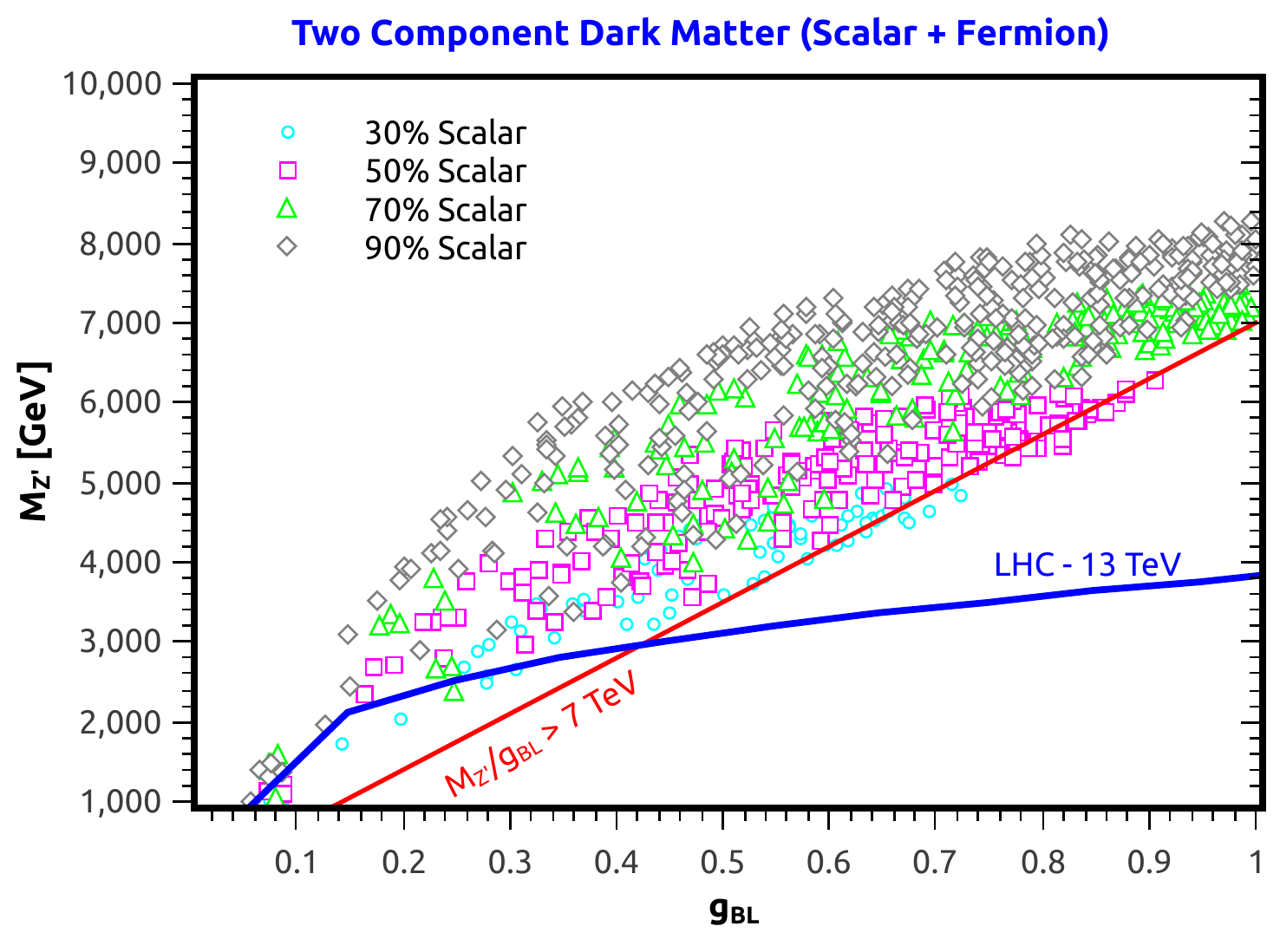}
\caption{Scan of the parameter space, in which a two-component dark matter scenario can be successfully realized and account for the entire dark matter of the Universe in agreement with direct detection limits. We have superimposed limits from the LHC (blue curve) and LEP (red curves). The points with different shapes represent different scalar dark matter contributions to the overall dark matter abundance. Blue circles represent the scenario where the scalar makes up for 30\% of the total abundance; pink squares correspond to 50\% of the total abundance; green triangles correspond to 70\% of the total abundance; and gray diamonds correspond to 90\% of the total abundance.}
\label{fig:11}
\end{figure}

In Fig.~\ref{fig:11} we investigate the possibility of having two-component dark matter (fermion plus scalar) making up the total abundance. All the points are consistent with direct detection limits. As an example, we fix $n=1/3$ for the fermion and $n=1$ for the scalar and let the dark matter mass free. A scan in the plane $M_{Z^{\prime}}$ vs.\ $g_{BL}$ is performed looking for regions where $\Omega h^2=0.11-0.12$. 
We have learned in the previous sections that scalar dark matter is more constrained than the Dirac fermion case, and for this reason we chose to exhibit several regimes for the two component dark matter based on the scalar abundance.
Blue circles represent the scenario where the scalar makes up for 30\% of the total abundance; pink squares correspond to 50\% of the total abundance; green triangles correspond to 70\% of the total abundance; and gray diamonds correspond to 90\% of the total abundance.  Limits from the LHC (blue curve) and LEP (red curves) are also shown.

Notice that there are large regions of parameter space, where a two-WIMP dark matter scenario is realized within a well motivated theory.  
Since the interactions that govern the scalar dark matter abundance are not very efficient, the scalar-dominated regime easily overcloses the Universe. The way out is to use sufficiently large gauge couplings and live near the $Z^{\prime}$ resonance region, enhancing the annihilation cross section and consequently bringing down the abundance to the proper value. Basically, all points in Fig.~\ref{fig:11} are in the neighbourhood of the resonance, except those for $g_{BL} \sim 1$, where one can obtain the right relic density while being slightly away from the resonance. This feature was observed in Figs.~\ref{fig:xsection}-\ref{fig:xsectionscalar}.

The points representing different regimes overlap, because we are scanning over the dark matter mass, which largely changes the abundance of the Dirac fermion dark matter. Therefore, for the same $g_{BL}$ one might have different abundances for the scalar and fermion fields, which explains the overlapping. In summary, Fig.~\ref{fig:11} shows a UV complete realization of a two component dark matter scenario.

\section{Conclusions}
Supplementing the SM with an extra $\mathrm{U}(1)_{B-L}$ gauge symmetry is an appealing possibility. In this paper, we studied the dark matter phenomenology of simplified models exhibiting such a gauge symmetry and in particular the possibilities of having Dirac fermion as well as scalar dark matter with and without broken B-L symmetry. In this context, we determined the impact of constraints coming from indirect and direct detection experiments as well as collider limits. Bounds from LUX2015, LUX2016 and projected bounds from XENON1T have been considered along with the famous LEP limit. In addition, we re-interpreted dilepton searches from the LHC at 13 TeV and extracted competitive limits for the model. 

While XENON1T projected bounds have a very good potential to exclude most of if not all the parameter space for scalar dark matter, we found that Dirac fermion dark matter would still be viable in a larger region of the parameter space. Interestingly, it was shown that the LHC limits that were extracted from dilepton production are already better than the LEP bounds for small gauge couplings. Finally, we also considered a mixed dark matter scenario, in which the relic abundance is realized as a combination of both fermion and scalar dark matter. In this case, numerous points satisfying the required relic density, collider, direct and indirect dark matter constraints were found, showing that a minimal and successful two component dark matter model is realized.


\section*{Acknowledments}

The work of M.K.\ was supported by the BMBF under contract 05H15PMCCA
and by the Helmholtz Alliance for Astroparticle Physics (HAP).
The work of F.L.\ was partially supported by the U.S.\ Department of Energy
under Grant No.\ DE-SC0010129. The authors thank Carlos Yaguna, Alexandre Alves, and Juri Smirnov for discussions.

\bibliographystyle{JHEP}
\bibliography{darkmatter}

\end{document}